\ifdefined\pdfminorversion\pdfminorversion=7\relax\fi
\documentclass[sigconf,nonacm]{acmart}
\setcopyright{none}

\newcommand{\artifactavailabilityurl}{https://github.com/Concyclics/MemForest}

\usepackage{float}
\usepackage[ruled,vlined,linesnumbered]{algorithm2e}
\usepackage{algpseudocode}
\usepackage{booktabs}
\usepackage{graphicx}
\usepackage{amsmath}
\usepackage{enumitem}
\usepackage[table]{xcolor}
\usepackage[most]{tcolorbox}
\usepackage{tabularx}
\usepackage{multirow}
\usepackage[normalem]{ulem}
\usepackage{tikz}
\usetikzlibrary{arrows.meta,positioning,fit,calc,backgrounds}

\newcommand{\artifactblob}[2]{\href{https://github.com/Concyclics/MemForest/blob/main/#1}{#2}}
\newcommand{\artifacttree}[2]{\href{https://github.com/Concyclics/MemForest/tree/main/#1}{#2}}
\newcommand{\fullonly}[1]{#1}
\newtcolorbox{promptbox}[1][]{
  enhanced,
  breakable,
  colback=white,
  colframe=black,
  boxrule=0.4pt,
  arc=1pt,
  outer arc=1pt,
  left=4pt,
  right=4pt,
  top=4pt,
  bottom=4pt,
  boxsep=0pt,
  before skip=6pt,
  after skip=6pt,
  fonttitle=\bfseries,
  title=#1
}

\usepackage{tcolorbox}

\definecolor{ChallengeBlue}{RGB}{27,58,91}
\definecolor{ChallengeBackground}{RGB}{242,247,251}

\newtcolorbox{challenge}[2]{
  colback=ChallengeBackground,
  colframe=ChallengeBlue,
  colbacktitle=ChallengeBlue,
  coltitle=white,
  fonttitle=\bfseries,
  title={Challenge #1: #2},
  arc=0pt,
  outer arc=0pt,
  boxrule=0.5pt,
  boxsep=0pt,
  left=6pt,
  right=6pt,
  top=5pt,
  bottom=5pt,
  toptitle=4pt,
  bottomtitle=4pt,
  before skip=7pt,
  after skip=7pt
}

\definecolor{ContributionGreen}{RGB}{32,91,67}
\definecolor{ContributionBackground}{RGB}{242,248,244}

\newtcolorbox{contribution}[2]{
  colback=ContributionBackground,
  colframe=ContributionGreen,
  colbacktitle=ContributionGreen,
  coltitle=white,
  fonttitle=\bfseries,
  title={Contribution #1: #2},
  arc=0pt,
  outer arc=0pt,
  boxrule=0.5pt,
  boxsep=0pt,
  left=6pt,
  right=6pt,
  top=5pt,
  bottom=5pt,
  toptitle=4pt,
  bottomtitle=4pt,
  before skip=7pt,
  after skip=7pt
}

\newtcolorbox{finding}[2]{
  colback=ContributionBackground,
  colframe=ContributionGreen,
  colbacktitle=ContributionGreen,
  coltitle=white,
  fonttitle=\bfseries,
  title={Finding #1: #2},
  arc=0pt,
  outer arc=0pt,
  boxrule=0.5pt,
  boxsep=0pt,
  left=6pt,
  right=6pt,
  top=4pt,
  bottom=4pt,
  toptitle=4pt,
  bottomtitle=4pt,
  before skip=6pt,
  after skip=6pt
}

\newtheorem{proposition}{Proposition}

\newcommand{\RthreeUnitsLME}{12}
\newcommand{\RthreeUnitsLoCoMo}{5}

\newcommand{\MemZeroMaintTime}{91.1\%}\newcommand{\MemZeroMaintOut}{91.4\%}
\newcommand{\MemOSMaintTime}{94.0\%}\newcommand{\MemOSMaintOut}{94.2\%}
\newcommand{\MFMaintTime}{53.7\%}\newcommand{\MFMaintOut}{27.6\%}
\newcommand{\MFExtractTime}{46.3\%}

\begin{document}

\title{MemForest: Write-Efficient Agent Memory with Decoupled Extraction and Hierarchical View Maintenance}
\settopmatter{authorsperrow=5,printacmref=false,printccs=false}

\author{Han Chen}
\affiliation{\institution{National University of Singapore}\city{Singapore}\country{Singapore}}
\email{chenhan@u.nus.edu}
\author{Zining Zhang}
\affiliation{\institution{National University of Singapore}\city{Singapore}\country{Singapore}}
\email{zzn@nus.edu.sg}
\author{Wenqi Pei}
\affiliation{\institution{National University of Singapore}\city{Singapore}\country{Singapore}}
\email{wenqi\_pei@u.\allowbreak nus.edu}
\author{Bingsheng He}
\affiliation{\institution{National University of Singapore}\city{Singapore}\country{Singapore}}
\email{dcsheb@nus.edu.sg}

\author{Ming Wu}
\affiliation{\institution{Zero Gravity Labs}\country{United States}}
\email{ming@0g.ai}
\author{Jason Zeng}
\affiliation{\institution{Zero Gravity Labs}\country{United States}}
\email{jason@0g.ai}
\author{Michael Heinrich}
\affiliation{\institution{Zero Gravity Labs}\country{United States}}
\email{michael@0g.ai}
\author{Wei Wu}
\affiliation{\institution{Zero Gravity Labs}\country{United States}}
\email{wei@0g.ai}
\author{Hongbao Zhang}
\affiliation{\institution{Zero Gravity Labs}\country{United States}}
\email{peter@0g.ai}
\author{Sen Gao}
\authornote{Corresponding author.}
\affiliation{\institution{National University of Singapore}\city{Singapore}\country{Singapore}}
\email{sen@u.nus.edu}

\renewcommand\shortauthors{Chen et al.}

\begin{abstract}
Long-lived LLM agents maintain semantic views, such as fact records, profiles,
and summaries, over a growing interaction history. Maintaining these views is
a principal cost of agent memory: ingesting one conversation takes 15--53
minutes with leading memory systems, and most of this time lies on chains of
dependent LLM calls that additional LLM capacity barely shortens. Because an
LLM-generated view admits no delta rule, existing designs either
serialize writes around revising views or give up fresh, reusable views. We
present \textbf{MemForest}, which treats every view as a result derived
from durable facts and recomputes it only from its children in a
time-ordered tree. The dependency depth of view refresh then follows tree
height rather than history length. Facts are extracted independently of
existing memory, shared ancestors are refreshed once per batch, and
independent views refresh in parallel. At query time, links from facts to
trees recover details that compact views omit, and materialized interval views
are reused for time-range aggregation. Against six released systems, MemForest
builds memory \(4.7\times\) faster than EverMemOS on LongMemEval-S and
\(8.8\times\) faster on LoCoMo. With Qwen3-30B, it matches EverMemOS on LoCoMo
(84.1\%) and attains the highest LongMemEval-S accuracy (81.8\%).

\end{abstract}

\keywords{LLM agents, agent memory, materialized views, incremental view maintenance, holistic aggregates, temporal aggregation, decoupled extraction}

\maketitle

\begingroup
\renewcommand\thefootnote{}%
\footnotetext{\noindent\textbf{Code:} \url{\artifactavailabilityurl}}%
\endgroup

\section{Introduction}
\label{sec:intro}



Large language model (LLM) agents increasingly operate over interactions that
span days, weeks, or months
\cite{park2023generative,packer2023memgpt,zhong2024memorybank,tang2026llm}.
Such an agent must incorporate new preferences, commitments, and experiences
as they occur, and must answer ad hoc queries over the accumulated history
\cite{zhang2025survey,lightmem}. Agent memory systems therefore maintain
\emph{semantic views} over the history: fact records, user profiles, episode
summaries, or graph summaries that an LLM generates from the dialogue
\cite{mem0,memoryos,lightmem,evermemos,zep}. Like materialized views in a
database, these views consolidate many observations ahead of time. A query
about the current state of an entity reads one consolidated record instead of
reconciling scattered and possibly outdated facts, and an aggregate request,
such as a profile or a summary of a period, reuses work that would otherwise be
repeated on every read. Precomputed hierarchical summaries, for example,
answer global questions with 9--43\(\times\) fewer context tokens than
summarizing the source text at query time \cite{edge2024local}.

\begin{figure}[t]
\centering
\includegraphics[width=\columnwidth]{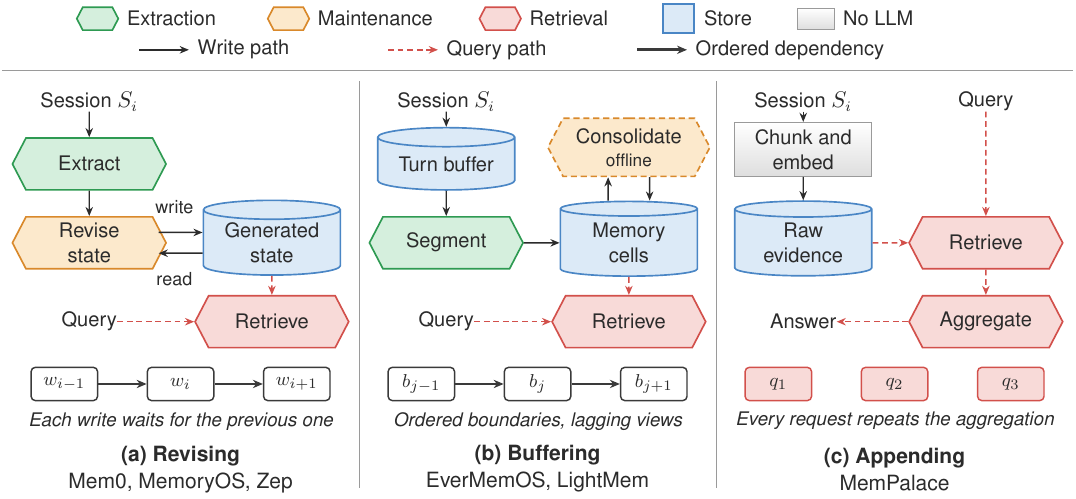}
\caption{Write and query paths of the three existing design strategies.}
\label{fig:strategies}
\end{figure}

View maintenance is a principal cost of agent memory. Before an observation
becomes visible to queries, it passes through a sequence of dependent LLM
stages, such as extraction, reconciliation against existing records, and
summarization, where each stage consumes the output of the previous one. With
Qwen3-30B-A3B served on one NVIDIA H100 GPU, ingesting a single LoCoMo
conversation of about 20K tokens takes 15 minutes with EverMemOS
\cite{evermemos}, 19 with Mem0 \cite{mem0}, 46 with MemoryOS \cite{memoryos},
and 53 with Zep \cite{zep}. This visibility latency determines how stale memory
is. Until a write completes, queries read a memory that lacks the latest
session, so an agent cannot rely on what it has just been told. Serving more
users in parallel or provisioning more GPUs raises aggregate throughput, but it
does not shorten the write latency of one history, because that latency arises
from a chain of dependent LLM calls within the history. The quantity to reduce
is therefore the length of this dependency chain rather than the total work.

Existing systems follow three design strategies (Figure~\ref{fig:strategies}),
and each sacrifices write parallelism, view visibility, or view reuse.
\emph{Revising} systems update generated state in place as evidence arrives.
Mem0 reconciles each new fact against retrieved records, MemoryOS rewrites
pages and a user profile, and Zep resolves new entities and edges against its
graph. Because each update reads records that preceding updates wrote, writes
form a dependency chain. In our traces, even unbounded parallelism shortens a
build by at most \(1.6\times\), and maintenance accounts for
\MemZeroMaintTime{} and \MemOSMaintTime{} of build-time LLM latency in Mem0 and
MemoryOS. Revision also determines which evidence survives. A rewritten view
preserves earlier detail only by regenerating it, and once the view reaches its
length limit, new detail displaces old. In MemoryOS, 315 of 324 profile
rewrites reach the 2,000-token output limit, and for 21\% of questions the
evidence is stored but never reaches the answer context. \emph{Buffering}
systems accumulate turns before forming memory, which relocates the dependency
rather than removing it. EverMemOS segments the stream at boundaries that
depend on its earlier decisions, and LightMem \cite{lightmem} consolidates
frozen snapshots offline, so its views lag behind the conversation.
\emph{Appending} systems such as MemPalace \cite{mempalace} store and index raw
evidence without generating views. Their writes are inexpensive, but every
aggregate request must be recomputed at query time, which moves the LLM work
onto query latency and repeats it for every request.

The underlying difficulty is that LLM-generated views admit no incremental
maintenance rule. Database systems keep materialized views current by
propagating deltas \cite{gupta1995maintenance}, which requires a view to be
computable from its previous value and the change. An LLM summary instead
behaves like a holistic aggregate \cite{gray1997data}: no rule derives the new
text from the old text and a delta. A system can revise the previous text,
which chains every write to its predecessor, or regenerate the view from all
evidence in its scope, whose input grows with the history and soon exceeds what
one summary call can process accurately. Neither option confines the work of a
write to the data it changes.

MemForest takes a different approach. It stores extracted facts as durable
evidence, separate from the views, and treats every view as a result derived
from that evidence rather than as mutable state. Views are organized into
time-ordered trees, called \emph{MemTrees}, and each internal view is
recomputed from the current views of its children, never from its own previous
text. Prior hierarchical memories either build their trees once offline
\cite{sarthi2024raptor} or merge each new item into the previous text of every
ancestor \cite{rezazadehisolated}. Since each refresh reads a bounded number of children, a write
regenerates only the views on one leaf-to-root path per affected tree, and the
dependency depth of view refresh follows tree height rather than history
length. A view therefore never has to carry every detail of its scope. Facts
remain in the store and stay reachable from the trees, so views can remain
short while the evidence they omit stays available to queries.

However, making it efficient raises three challenges.

\noindent\textbf{C1: Write parallelism under state dependence.} Extraction does
not depend on existing memory, whereas deduplicating and committing new facts
must observe the facts already stored. MemForest extracts candidates from short
dialogue cells that see only their own text, and then canonicalizes them
against the fact store before any view is touched. Independent work thus runs
concurrently, and only the decisions that read memory state remain ordered.

\noindent\textbf{C2: Refresh cost at scale.} A single recomputation is local,
but the writes of a batch share ancestors, and refreshing once per write
repeats identical LLM calls. MemForest marks the affected paths, refreshes each
shared ancestor once per batch, and processes the dirty nodes of all trees
level by level, so independent nodes refresh in parallel. The tree fan-out
keeps each refresh within what one summary call handles accurately.

\noindent\textbf{C3: Detail and aggregation under compact views.} The shorter a
view, the more often a point query needs a detail that the view omits, whereas
aggregate requests should reuse views rather than reread facts. MemForest links
each fact to the trees that contain it, so a matching fact directs the query to
the relevant tree even when the root omits the detail, and embedding-guided
browsing collects the supporting evidence. A time-range request is answered
from a small set of interval views already materialized in the tree, and the
size of this set does not grow with the range.

This paper makes four contributions:
\begin{itemize}
\item \textbf{Agent memory as view maintenance.} We formulate agent memory as
the maintenance of derived views over durable facts. A measurement study of
four released systems shows how revising views serializes writes, regenerates
content that views already hold, and leaves stored evidence out of answers.
\item \textbf{A write path with refresh depth bounded by tree height.}
MemForest extracts facts from independent cells and recomputes each affected
view from its children, refreshing shared ancestors once per batch, level by
level, across trees.
\item \textbf{Fact-guided navigation.} Fact-to-tree links keep detail reachable
under compact views, and range covers answer aggregates from a bounded set of
materialized views.
\item \textbf{Extensive evaluation.} Against six released systems on two
benchmarks and three LLM backbones, MemForest builds memory 4.7--8.8\(\times\)
faster than EverMemOS while matching or exceeding the accuracy of the best
baseline. Controlled studies isolate the effect of each mechanism.
\end{itemize}

\section{Problem and Motivation}
\label{sec:background}

\begin{figure*}[t]
\centering
\includegraphics[width=\textwidth]{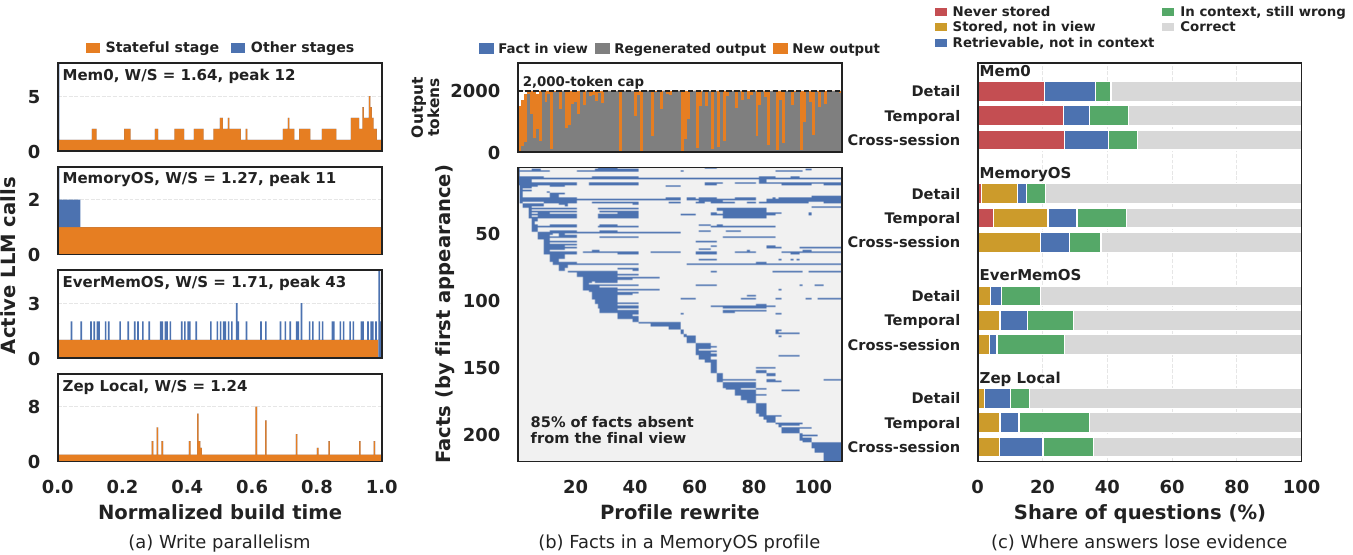}
\caption{Write paths and failures of existing systems.}
\label{fig:motivation}
\end{figure*}

\subsection{Problem}
\label{subsec:workload_model}
\label{subsec:temporal_scope}

An agent observes a time-ordered stream of sessions
\(\mathcal{D}_T=(S_1,\ldots,S_T)\), each a bounded sequence of timestamped
turns. A \emph{write} ingests one session. A \emph{semantic view} is text that
an LLM generates over the evidence in a \emph{scope}, such as a session or an
entity. A write takes effect when queries can observe the updated evidence and
views, and the time until then is what an agent waits for.

Queries come in two kinds. A \emph{point lookup} is answered by a few facts and
gains little from a view. An \emph{aggregate request} asks for a summary over a
scope or a time range of it, and without a view it must read and combine every
relevant fact at query time. At eight-session ranges, reusing materialized
views is \(2.3\times\) faster than map-reducing the fact log
(Figure~\ref{fig:range}). On LoCoMo, a maintained entity view also costs fewer
tokens than map-reduce once it has served about 12 eight-session summaries, or
3 summaries of the entity's full history. Views are worth maintaining only if refreshing them is cheap, and current
systems do not make it cheap.

\subsection{Existing Designs}
\label{subsec:design_space}

\begin{table}[t]
\centering
\scriptsize
\caption{Write-path LLM dependencies in existing memory systems.
Ordered: a call observes state committed by an earlier call. Independent: calls
can run concurrently. None: the stage issues no LLM call.}
\label{tab:design_space}
\label{tab:write_path_critical_path}
\setlength{\tabcolsep}{2.4pt}
\renewcommand{\arraystretch}{1.12}
\begin{tabular}{@{}llcccc@{}}
\toprule
& & \multicolumn{2}{c}{\textbf{LLM dependency}} & \textbf{Revises} & \textbf{View} \\
\cmidrule(lr){3-4}
\textbf{Strategy} & \textbf{System} & \textbf{Extraction} & \textbf{Maintenance}
& \textbf{state} & \textbf{visibility} \\
\midrule
\multirow{3}{*}{Revising} & Mem0~\cite{mem0} & Independent & Ordered & \checkmark & Immediate \\
& MemoryOS~\cite{memoryos} & None & Ordered & \checkmark & Immediate \\
& Zep Local~\cite{zep} & Ordered & Ordered & \checkmark & Immediate \\
\midrule
\multirow{2}{*}{Buffering} & EverMemOS~\cite{evermemos} & Ordered & None & \(\times\) & Immediate \\
& LightMem~\cite{lightmem} & Independent & Independent & \checkmark & Delayed \\
\midrule
Appending & MemPalace~\cite{mempalace} & None & None & \(\times\) & No views \\
\bottomrule
\end{tabular}
\end{table}

Table~\ref{tab:design_space} classifies existing designs by the LLM dependencies
of their write paths. Revising systems maintain the view of a scope
\(\sigma\) by editing its previous text. Let \(s_\sigma^{(i)}\) be the view after
the \(i\)-th update to \(\sigma\) and \(\Delta E_\sigma^{(i)}\) the evidence that
update adds. An LLM call then produces
\begin{equation}
s_\sigma^{(i)}=\textsc{Update}\bigl(s_\sigma^{(i-1)},\Delta E_\sigma^{(i)}\bigr),
\label{eq:state_transition_scope}
\end{equation}
so consecutive writes to a scope cannot be separated. The alternative is to
recompute from the scope's full evidence \(E_\sigma\),
\(s_\sigma=\textsc{Summarize}(E_\sigma)\), but this input grows with
\(|E_\sigma|\) and a single call loses detail once it exceeds what one summary
can hold. In our measurements, per-fact recall stays at or above 0.95 only up to
about 16 inputs (Figure~\ref{fig:fanout}a). Buffering designs move
the dependency to segmentation or offline, and appending designs remove it by
giving up reusable views.
\fullonly{Appendix~\ref{app:baseline-write-path-analysis} gives each system's
stage-level schedule.}

\subsection{Measurement Study}
\label{subsec:motivation_study}
\label{subsec:existing_limitations}

Table~\ref{tab:design_space} shows where each design places its LLM
dependencies, but not what they cost. We run three experiments on Mem0,
MemoryOS, EverMemOS, and Zep Local, each answering one question the table
leaves open: whether writes wait because of real dependencies, what a revised
view costs as the history grows, and where the evidence of a failed answer
ends up (Figure~\ref{fig:motivation}).

\noindent\textit{\textbf{Dependency chains.}}
Maintenance takes \MemZeroMaintTime{} and \MemOSMaintTime{} of build-time LLM
latency in Mem0 and MemoryOS. Released systems run their writes one at a time,
so measured build time cannot tell whether this waiting comes from the data or
from how the code is written. We therefore record every LLM call together with
the memory objects it reads and writes, and replay each trace with unlimited
concurrency, issuing every call as soon as the objects it reads exist. The
replay gives the parallelism that the dependencies themselves allow. Even so,
every system averages fewer than two active calls, that is, its total work
\(W\) is less than twice its critical path \(S\) (Figure~\ref{fig:motivation}a),
and 128 concurrent slots shorten a build by at
most 1.6\(\times\). The cause is the stateful stage, such as reconciliation in
Mem0 or profile rewriting in MemoryOS, which forms one chain that every write
joins. The waiting is inherent in revising shared state, and more LLM capacity
cannot remove it.

\noindent\textit{\textbf{View regeneration.}}
A view revised by Eq.~\ref{eq:state_transition_scope} keeps old detail only by
writing it out again. To measure this cost, we follow every version of a
MemoryOS profile, split each version into atomic facts, and mark which facts it
contains. We also split each rewrite's output into new and regenerated tokens,
and repeat the run with larger output caps. Most of each rewrite copies the
previous version, and 77.5\% of the output tokens are regenerated. The profile
soon reaches its 2,000-token cap, after which new facts displace old ones, and
87\% of the facts that ever enter it are absent from its final version
(Figure~\ref{fig:motivation}b). Raising the cap keeps more facts but lengthens
every rewrite, and 93\% of an 8,000-token rewrite is
regenerated\fullonly{ (Figure~\ref{fig:profile-cap})}. A revised view must
grow with the history or drop detail.

\noindent\textit{\textbf{Evidence delivery.}}
A wrong answer can come from a fact that was never stored, a view that omits
it, or a retrieval that misses it, and each calls for a different fix.
End-to-end accuracy cannot separate these cases. For 600 questions per system,
we locate the supporting evidence along the path from storage through views and
retrieval to the answer context, and assign each question to the first stage
that drops it. Failed answers often have their evidence in memory
(Figure~\ref{fig:motivation}c). Mem0 keeps only the facts it extracts, with no view layer, and
never stores the evidence for 23.8\% of questions. MemoryOS stores it, but for 21\%
of questions neither its profile nor retrieval delivers it to the answer. Such
questions are common, since temporal, knowledge-update, and multi-session
questions make up about two thirds of LongMemEval-S \cite{longmemeval}.

All three problems share one cause: these systems produce the memory state that
queries read by revising earlier state. Revision puts every write on one chain,
makes each refresh regenerate what the view already holds, and lets the
revision decide which evidence survives. A memory that keeps views current must
therefore separate durable evidence from the views derived from it. Facts
should be stored apart from views, a write should wait only for the decisions that
read memory state, a view should be refreshed from bounded current input rather
than from its own previous text, and a query should be able to follow a view
back to the facts it omits.

\section{MemForest Overview}
\label{sec:system}

\begin{figure}[t]
    \centering
    \includegraphics[width=\columnwidth]{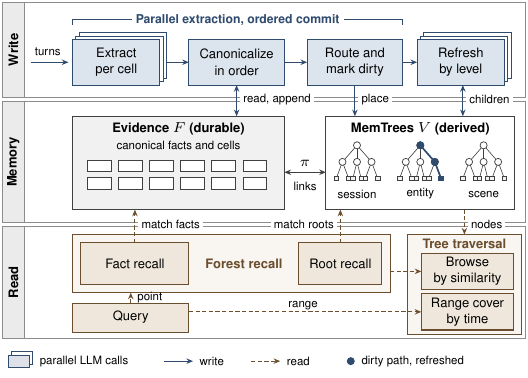}
    \caption{MemForest architecture.}
    \label{fig:memforest_overview}
\end{figure}

\subsection{Data Model and Architecture}
\label{subsec:data_model}

MemForest keeps evidence and views apart
(Figure~\ref{fig:memforest_overview}). It represents an agent's memory as a
forest \(\mathcal{M}=(F,\pi,V)\) with one time-ordered tree, called a MemTree,
per scope. A scope \(\sigma\) is a session, an entity, or a topical scene. The
fact store \(F\) holds canonical facts and the short dialogue \emph{cells} they
are extracted from, and each fact keeps its source cell, source time, and any
event time resolved from the text. The \emph{placements} \(\pi\) record the
leaves at which a fact or cell appears in each scope. \(V\) holds the views of
tree nodes. A leaf's view is its fact or cell text, and the view \(s_v\) of an
internal node \(v\) is generated by an LLM summarizer \(g\) from the views of its
children \(\mathrm{Ch}(v)\). Views, node embeddings, and index rows are
\emph{derived artifacts} because MemForest can regenerate them from \(F\) and
\(\pi\).

A write first extracts fact candidates from each cell without reading existing
memory. Canonicalization then integrates the candidates into \(F\), routing
places them into scopes, and affected views are recomputed from their children.
We call all work after extraction \emph{maintenance}, because it reads or updates
existing state. A query uses the same placements in the other direction, from
matching facts and views back to the evidence in the trees.

\subsection{Local Recomputation and Fact-Guided Navigation}
\label{subsec:design_insight}

\begin{figure}[t]
    \centering
    \includegraphics[width=\columnwidth]{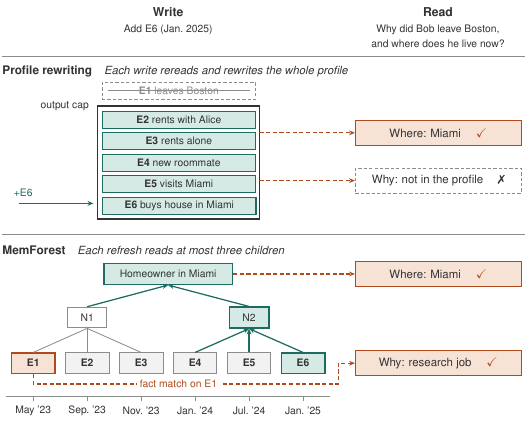}
    \caption{Running example: writing a new evidence item and
    answering a question under profile rewriting and under MemForest. Green marks
    what the write changes, and red what the read uses.}
    \label{fig:example}
\end{figure}

\noindent\textit{\textbf{Running example.}}
\label{subsec:temporal_memory_problem}
\label{subsec:problem_statement}
Consider the constructed history illustrated in
Figure~\ref{fig:example}. Bob leaves Boston
for a research job in Davis in \textbf{May 2023}, rents with Alice in
\textbf{Sep.\ 2023}, rents alone in \textbf{Nov.\ 2023}, has a new roommate in
\textbf{Jan.\ 2024}, visits Miami in \textbf{Jul.\ 2024}, and moves to Miami and
buys a house there in \textbf{Jan.\ 2025}. We name these facts E1--E6. A compact
view, \emph{Bob is a homeowner in Miami}, answers a current-city question but
omits why he left Boston, and a view that keeps those details must reproduce
them whenever it is revised.

Figure~\ref{fig:example} contrasts the two maintenance styles. A profile in
the style of MemoryOS \cite{memoryos} is rewritten as a whole when new evidence
arrives, so adding E6 regenerates every line. Once the profile reaches its
output cap, the new line also pushes out the oldest one, E1.

MemForest instead recomputes a view from its current children. When a write
changes the leaves below an internal node \(v\), the node becomes \emph{dirty},
and MemForest computes
\begin{equation}
s_v^{\mathrm{new}}=
g\bigl(\langle s_c^{\mathrm{current}}\rangle_{c\in\mathrm{Ch}(v)}\bigr).
\label{eq:local_view_recomputation}
\end{equation}
Unlike Eq.~\ref{eq:state_transition_scope}, the refresh never reads the
previous view of \(v\), and it reads at most \(f\) current children rather than
all evidence in the scope, where the fan-out \(f\) bounds the children of a node.
In Bob's tree, internal node N1 covers E1--E3 and N2 covers E4--E6. Adding E6
refreshes N2 from E4--E6 and then the root from N1 and N2, while N1 and all
five earlier facts are reused unchanged.

The two styles differ in what one write puts into the LLM's context. A profile
rewrite reads the whole previous profile and emits it again, so its cost grows
with the detail the profile keeps until the cap forces it to drop detail
instead. A MemForest refresh reads at most \(f\) child views and emits one short
view, and a write needs one such refresh per tree level. MemForest caps its
views too, at a word target far below the profile cap, but capping a view drops
no evidence. Facts stay in \(F\), and the placements \(\pi\) lead a query from a
matching fact back to its tree.

For ``Why did Bob leave Boston, and where does he live now?'', the root view
answers where but omits the research-job reason. E1 matches the question and
leads to Bob's tree, so the reason stays reachable, and this is why views can
stay short. Each internal node also covers a time
span, so a request for Bob's housing between May 2023 and Jan.\ 2024 reads only the
view of N1 and the leaf E4 instead of summarizing every fact in that
period. View refresh changes derived artifacts but never replaces accepted
evidence. Sections~\ref{sec:write-path}--\ref{sec:lifecycle} specify the write
path, view maintenance, the query path, and the lifecycle.

\section{Decoupled Write Path}
\label{sec:write-path}
\label{sec:impl}
\label{sec:impl:extract}

A write first extracts fact candidates, then canonicalizes and places them
before maintaining views. Extraction reads only the incoming cells. We use the running example of
Section~\ref{subsec:temporal_memory_problem} throughout.

\noindent\textit{\textbf{Cells and extraction.}}
MemForest partitions a session \(S_t\) into cells of at most \(b\) consecutive
turns,
\begin{equation}
\mathcal{C}(S_t)=\{c_{t,j}\}_{j=1}^{\lceil n_t/b\rceil},\ \
c_{t,j}=(u_{t,(j-1)b+1},\ldots,u_{t,\min(jb,n_t)}),
\end{equation}
and also closes a cell at a long time gap or a size limit. We use \(b=2\):
two turns keep enough context to resolve local references while keeping each
call short (Figure~\ref{fig:chunks}). An extraction call sees only its cell
and the cell's timestamp, never stored facts or views, so the calls of a
session, and of several sessions, are mutually independent. They run in
\(\lceil|\mathcal{C}|/P\rceil\) \emph{waves}, sets of independent requests
issued together, at concurrency \(P=128\). Each call returns
fact candidates with source spans, origin and speaker role, entities, topics,
confidence, and an event time when the text states one; a candidate without
an event time inherits its cell's time.

\noindent\textit{\textbf{Canonicalization.}}
Canonicalization reads the existing fact store \(F\) and integrates the
extracted candidates before structural and view maintenance. Candidates whose normalized text matches an
existing fact merge into it. Each remaining candidate retrieves its nearest
facts by embedding similarity, and an LLM judge decides equivalence only for
pairs above a high similarity threshold (at most four per candidate). The
judge rejects pairs that differ in date, occurrence, entity, quantity, or
role, so an equivalent candidate is appended to the canonical fact as another
occurrence, while an update such as ``rents alone'' after ``rents with Alice''
becomes a new, separately time-stamped fact. Canonicalization never rewrites a
fact's text and never reads a view; it issues 19--47 judge calls per evaluated
history. We count it as maintenance in
all measurements.

\noindent\textit{\textbf{Routing and placements.}}
Each canonical fact is placed into three types of scope without further LLM
calls. (1)~In the \emph{session scope}, every session has a tree whose leaves
are its cells, and a cell leaf points to the canonical facts extracted from
it. (2)~In the \emph{entity scope}, the user tree receives every user-origin
fact, and an entity receives its own tree once a specific name is mentioned
by at least three facts from at least two sessions. (3)~In the \emph{scene
scope}, a fact is routed by embedding similarity, with a small recency bonus,
to topical clusters of the same domain; it joins the best one or two clusters
or spawns a new one, and clusters are capped at 60 facts. The placements of fact \(r\) are thus its
cell leaves, its user-tree leaf, one leaf per active entity, and one or two
scene leaves. A 300-fact audit finds entity routing precise but incomplete:
124/127 active entity placements are semantically valid, but only 9/127 are
complete, i.e., include every expected entity placement. Entity trees are
therefore a precision-oriented overlay; a fact missing from an entity tree
remains reachable through its session and scene trees and through
fact-to-tree recall (Section~\ref{sec:query}). Routing produces the records
\begin{equation}
\mathcal{R}_t=\{(\sigma,r)\mid r\in\mathcal{F}_t,\ \sigma\in R(r)\},
\end{equation}
where \(\mathcal{F}_t\) are the facts touched by \(S_t\) and \(R(r)\) their
scopes, and hands them to view maintenance. The remaining thresholds (cell
closing at a 6-hour gap or 8,000 characters, canonicalization similarity 0.93
over 8 neighbors, scene spawn below 0.58, second assignment at 0.70, and merge
at 0.88) were not selected on benchmark accuracy and are listed in the
released configuration. Scene assignment is insensitive to
them: perturbing the merge threshold by \(\pm0.04\) keeps the mean assignment
Jaccard at or above 0.965.

\section{MemTree View Maintenance}
\label{sec:memtree}
\label{sec:impl:update}

\noindent\textit{\textbf{MemTree.}}
A MemTree \(\mathcal{T}_\sigma\) is an ordered tree over the placements of
scope \(\sigma\). Its leaves are ordered by
\((t_{\mathrm{start}},t_{\mathrm{end}},\mathit{id})\), where the times are a
fact's resolved event times (else its source time) or a cell's turn times.
Every internal node has between 2 and \(f\) children one level below, where
the fan-out \(f\) is 3 for session trees, 8 for entity trees, and 10 for scene
trees, and its time span is the union of its children's spans. We choose
\(f\) to keep the typical refresh near the high-recall region of
Figure~\ref{fig:fanout}a: a session leaf is a raw two-turn
cell several facts long, so session trees use \(f=3\), while entity and scene
trees hold single facts and use \(f=8\) and 10, near the root-recall peak of
Figure~\ref{fig:fanout}b. A leaf's view
is its payload verbatim, the fact text or the cell text, so leaves cost no LLM
call. An internal node's view is
\begin{equation}
s_v=g\bigl(\mathrm{hdr}(v),\langle s_c\rangle_{c\in\mathrm{Ch}(v)}\bigr),
\label{eq:node_view}
\end{equation}
where \(\mathrm{hdr}(v)\) holds the tree label, type, level, and time span.
The prompt requests \(W\) words at most (250 for entity trees, 200 otherwise);
\(W\) is a generation target. Leaf payloads remain verbatim. A view therefore depends only on its current children and never
on its own previous value. The structure invariants (a single parent-free
root, one payload per leaf, 2--\(f\) children per internal node, consistent
parent and child links, and a time-ordered leaf chain) are checked after every
build.

\noindent\textit{\textbf{Structural edits.}}
Insertion places a leaf at its time position in the leaf chain, so a fact
whose event time precedes newer facts lands in the middle of the tree rather
than at its end. An internal node that exceeds \(f\) children splits in half,
and splits propagate upward as in a B\(^+\)-tree; deletion borrows from or
merges with a sibling on underflow. Each edit marks the touched leaf and its
ancestors dirty and recomputes their time spans; no view is regenerated
during structural edits.

\begin{figure}[t]
    \centering
    \includegraphics[width=\columnwidth]{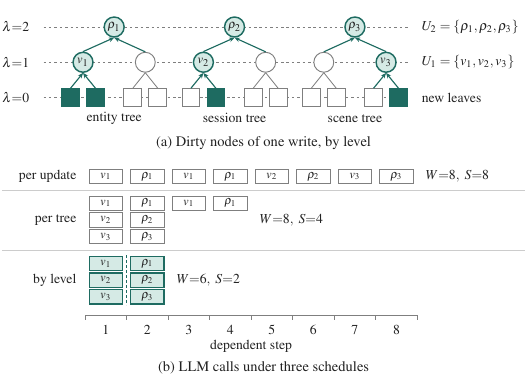}
    \caption{Coalesced refresh of one write.}
    \label{fig:coalesced}
\end{figure}

\noindent\textit{\textbf{Coalesced refresh.}}
Let \(D(u)\) be the internal nodes dirtied by update \(u\). Dirty nodes are
kept as sets, so a batch \(B\) refreshes
\begin{equation}
D(B)=\bigcup_{u\in B}D(u)
\label{eq:dirty_union}
\end{equation}
once each. Refresh proceeds by level: all dirty nodes of level \(\lambda\),
across all trees, form one pool \(U_\lambda\) that is issued at concurrency
\(P\), and level \(\lambda+1\) starts when level \(\lambda\) completes
(Figure~\ref{fig:coalesced} and Algorithm~\ref{alg:memtree_update}). In
Figure~\ref{fig:coalesced}, one write inserts four leaves into three trees, two
of them below the same node \(v_1\). Refreshing one update at a time issues
eight dependent calls (\(W{=}8\) calls in \(S{=}8\) dependent steps). Running the trees in parallel shortens the chain to four
steps but still refreshes \(v_1\) and \(\rho_1\) twice. Coalescing refreshes each
dirty node once, and issuing the pools level by level finishes in two steps,
one per level (\(W{=}6\), \(S{=}2\)). Node embeddings and root-index rows are
then recomputed for the refreshed nodes. A failed call leaves its node dirty
with its previous view, so the next refresh retries it. For example, sixteen
updates under four children of one root need five calls in two waves instead
of thirty-two, and adding E6 to Bob's tree refreshes only the parent of
E4--E6 and then the root (Figure~\ref{fig:example}). Batching skips intermediate view versions. The continuous-update experiment
in Section~\ref{sec:eval-maintenance} measures refresh savings and question
answering at shared, fully refreshed checkpoints. The \emph{dependency depth} of
a refresh is the number of LLM requests on its longest chain.

\begin{proposition}[Dependency depth]
\label{prop:depth}
Let \(h_\sigma\) be the height of \(\mathcal{T}_\sigma\). A successful refresh
of batch \(B\) has dependency depth at most \(\max_\sigma h_\sigma\).
A level-barrier schedule with at most \(P\) requests per wave uses
\(\sum_{\lambda\ge1}\lceil|U_\lambda|/P\rceil\) execution waves.
For \(n>1\) leaves, bottom-up packed batch construction gives
\(h=\lceil\log_f n\rceil\); the balanced 2--\(f\) child invariant gives
\(\lceil\log_f n\rceil\le h\le\lfloor\log_2 n\rfloor\).
\end{proposition}

\noindent\textit{Proof sketch.} A level-\(\lambda\) view reads only its
children at level \(\lambda-1\), and leaves need no LLM call. Thus the longest
chain has one summary call per internal level. Packed grouping uses
\(\lceil\log_f n\rceil\) levels; a balanced tree with 2--\(f\) children per
internal node has \(2^h\le n\le f^h\). \hfill\(\square\)

\begin{proposition}[Work]
\label{prop:work}
Refreshing \(B\) issues \(|D(B)|\le\sum_{u\in B}|D(u)|\) summary calls.
If \(\ell_c\) is the realized word length of child \(c\), a call for node
\(v\) reads \(\sum_{c\in\mathrm{Ch}(v)}\ell_c\) words plus its header.
Let \(L\) and \(S\) bound the realized leaf and internal-view lengths in the
batch. Each call reads at most \(f\max\{L,S\}\) child words and emits at
most \(S\) words. The number of child inputs is bounded by \(f\), independently
of the total evidence count in the scope.
\end{proposition}

This separates structural locality from text length: a refresh reads only
its immediate children, while the word target controls typical internal-view
length. With \(N\) extraction cells, \(D_\lambda\) dirty internal nodes per
level, and \(C_{\mathrm{can}}\) canonicalization waves, the write schedule is
\begin{equation}
C_{\mathrm{MF}}=\Bigl\lceil\frac{N}{P}\Bigr\rceil+C_{\mathrm{can}}
+\sum_{\lambda=1}^{H}\Bigl\lceil\frac{D_\lambda}{P}\Bigr\rceil
=O\Bigl(\frac{N+D}{P}+H+C_{\mathrm{can}}\Bigr),
\label{eq:mf_critical_path}
\end{equation}
where \(D=\sum_\lambda D_\lambda\) and \(H\) is the maximum height.
This counts the successful request schedule; additional attempts contribute
their own waves and tokens. Each fact--tree placement changes one leaf and
\(O(h_\sigma)\) ancestors; finding its time position is separate from this
structural-edit count. Across 250,118 trees of two complete 500-question Qwen
builds, the observed maximum height is 4.

\begin{algorithm}[t]
\caption{Coalesced MemTree maintenance for one write}
\label{alg:memtree_update}
\small
\DontPrintSemicolon
\SetKwInOut{Input}{Input}
\SetKwInOut{Output}{Output}
\Input{forest \(\mathcal{M}\), routed records \(\mathcal{R}_t\), concurrency \(P\)}
\Output{updated \(\mathcal{M}\) and unresolved dirty nodes \(D_{\mathrm{rem}}\)}
\ForEach{\((\sigma,r)\in\mathcal{R}_t\)}{
    \(\ell\gets\mathrm{InsertByTime}(\mathcal{T}_\sigma,r)\)
    \tcp*{split on overflow}
    \(\mathrm{MarkDirty}(\ell\) and its ancestors\()\)\;
}
\For{\(\lambda\gets1\) \KwTo \(\max_\sigma h_\sigma\)}{
    \(U_\lambda\gets\) dirty level-\(\lambda\) nodes of all trees
    \tcp*{deduplicated set}
    \ForEach{\(v\in U_\lambda\) \textbf{in parallel, at most \(P\)}}{
        \(s'\gets g(\mathrm{hdr}(v),\langle s_c\rangle_{c\in\mathrm{Ch}(v)})\)\;
        \lIf{\(s'\neq\bot\)}{\(s_v\gets s'\); \(\mathrm{MarkClean}(v)\)}
    }
    barrier: wait until all requests in \(U_\lambda\) complete\;
}
re-embed refreshed nodes; update index rows\;
\Return{\((\mathcal{M},\mathrm{DirtyNodes}(\mathcal{M}))\)}\;
\end{algorithm}

\section{Query Path}
\label{sec:query}
\label{sec:impl:retrieval}

\noindent\textit{\textbf{Forest recall.}}
Given a query \(q\), MemForest nominates trees through two channels.
Root recall returns the trees whose root views are most similar to \(q\),
\begin{equation}
C_{\mathrm{root}}(q)=\operatorname{TopK}_{T\in\mathcal{F}}\,
\mathrm{sim}\bigl(e(q),e(r_T)\bigr),
\label{eq:root_recall}
\end{equation}
where \(e(\cdot)\) is the embedding and \(r_T\) the root view of tree \(T\);
single-fact scene trees are not indexed. Fact-to-tree recall retrieves the
\(M\) facts most similar to \(q\) and maps each back to the trees that place
it,
\begin{equation}
\begin{gathered}
R_{\mathrm{fact}}(q)=\operatorname{Top}_M\,\mathrm{sim}\bigl(e(q),e(f)\bigr),\\
C_{\mathrm{fact}}(q)=\textstyle\bigcup_{f\in R_{\mathrm{fact}}(q)}\mathrm{Trees}(f),
\end{gathered}
\label{eq:fact_recall}
\end{equation}
and ranks these trees by their hits
\(h(q,T)=|\{f\in R_{\mathrm{fact}}(q):T\in\mathrm{Trees}(f)\}|\). The candidate
list keeps \(C_{\mathrm{root}}(q)\) in rank order, appends the top-\(K\)
fact-nominated trees not already present, and adds the user tree, so at most
\(2K+1\) trees are browsed; we use \(K=10\) and \(M=30\). All similarities are
cosine similarities over normalized Qwen3 embeddings, and the fact store, the
root index, and the node index use FAISS \texttt{IndexFlatIP} exact search. Fact-to-tree recall is what keeps detail reachable:
for ``why did Bob leave Boston,'' the research-job fact nominates Bob's tree
although its compact root omits the reason (Figure~\ref{fig:example}).

\noindent\textit{\textbf{Embedding-guided browsing.}}
Each candidate tree is browsed with the query \(q\), top-down over the node index with a
global beam of \(B=10\): at each level, MemForest scores the children of the
frontier by the cosine similarity between \(e(q)\) and their node
embeddings, materialized from internal-node views or leaf evidence
(Algorithm~\ref{alg:memtree_update}). If \(L\) contains the leaves already
settled, it keeps the best \(B-|L|\) candidates, settles newly reached leaves,
and stops once \(B\) leaves are settled.
Algorithm~\ref{alg:query} specifies this embedding-ranked traversal, where
\(\oplus\) concatenates lists in order and drops duplicates.
Descent scores only the children of the frontier; candidate trees are browsed
in parallel. Entity and scene leaves resolve to their facts and
session leaves to the canonical facts of their cell.

\noindent\textit{\textbf{Answer context.}}
Resolved facts are deduplicated and sorted by cosine similarity to the
query. The main answer context admits these facts from at most \(B\) leaves
per tree over at most \(2K+1\) trees, without further truncation. Each fact is
rendered with its time, up to two source occurrences, and up to two
alternative phrasings recorded by canonicalization, so the answer model sees
when a fact was stated and when it happened.
Section~\ref{sec:eval-same-pipeline} fixes the evidence budget
\(k_e\) at 10, 30, or 50 facts or 2,048 tokens to compare against a log.

\noindent\textit{\textbf{Range views.}}
An aggregate request over scope \(\sigma\) and range \([t_1,t_2]\) is answered
from the views already materialized in \(\mathcal{T}_\sigma\). Because leaves
are ordered by start time, the leaves whose start time lies in
\([t_1,t_2]\) form a contiguous run of the leaf chain.
\(\mathrm{Cover}(\mathcal{T}_\sigma,[t_1,t_2])\) returns the canonical
decomposition of that run, the maximal nodes all of whose leaves lie in it,
found by descending only into nodes that straddle a range boundary. One merge
call then summarizes the cover views in time order under the same word target
\(W\).

\begin{proposition}[Range cover]
\label{prop:cover}
For a MemTree of height \(h\) and fan-out at most \(f\), the cover of any range
has at most \(2(f-1)\,h\) nodes (plus the root if the range covers the whole
tree), independent of the number of facts in the range; the merge call reads
the cover text, whose length is the sum of its nodes' realized view lengths.
The singleton-tree case uses its one leaf. With realized leaf and internal-view
bounds \(L,S\), the cover contains at most
\(\max\{1,2(f-1)h\}\max\{L,S\}\) words before merge-prompt headers.
\end{proposition}

\noindent\textit{Proof sketch.} As for a segment tree, cover nodes at each
level below the root are siblings of the two root-to-boundary paths: at most
\(f-1\) on each side. \hfill\(\square\)

A log, by contrast, must read every fact in the range, or summarize it level
by level at query time; Section~\ref{sec:eval-same-pipeline} compares the two
on range-summary requests.

\begin{algorithm}[t]
\caption{Query: forest recall and embedding-guided browsing}
\label{alg:query}
\small
\DontPrintSemicolon
\SetKwInOut{Input}{Input}
\SetKwInOut{Output}{Output}
\Input{query \(q\), forest \(\mathcal{M}\), recall \(K,M\), beam \(B\)}
\Output{evidence facts \(E_q\)}
\(\mathcal{C}\gets\mathrm{TopK}_{\mathrm{root}}(q)\,\oplus\,
\mathrm{TopK}_{\mathrm{hits}}(\mathrm{FactsToTrees}(\mathrm{Top}_{M}\mathrm{Facts}(q)))
\,\oplus\,\{\mathcal{T}_{\mathrm{user}}\}\)\;
\(E_q\gets\emptyset\)\;
\ForEach{\(\mathcal{T}\in\mathcal{C}\) \textbf{in parallel}}{
    \(\Phi\gets\{\mathrm{root}(\mathcal{T})\}\)\;
    \(L\gets\mathrm{leaves}(\Phi)\); \(\Phi\gets\Phi\setminus L\)\;
    \While{\(\Phi\neq\emptyset\) \textbf{and} \(|L|<B\)}{
        \(X\gets\operatorname{Top}_{B-|L|,\,v\in\mathrm{Ch}(\Phi)}
        \mathrm{sim}(e(q),e(v))\)\;
        \(L\gets L\cup\mathrm{leaves}(X)\);
        \(\Phi\gets X\setminus\mathrm{leaves}(X)\)\;
    }
    \(E_q\gets E_q\cup\mathrm{Resolve}(L)\)\;
}
\Return{\(\mathrm{SortByCosine}(E_q,e(q))\)}\;
\end{algorithm}

\section{Lifecycle and Visibility}
\label{sec:lifecycle}
\label{sec:impl:maintenance}
\label{sec:impl:freshness}

\noindent\textit{\textbf{Lifecycle operations.}}
Lifecycle operations follow the write rule: edit facts and placements first,
then refresh only the views on affected paths. \emph{Delete} uses the session
registry to find the facts, cells, and leaves of a retracted session, removes
them with underflow repair, and refreshes only the invalidated ancestors;
deleting Bob's January 2025 session refreshes only the paths above the Miami
leaves. \emph{Merge} of forest \(\mathcal{M}_B\) into \(\mathcal{M}_A\) first
canonicalizes \(B\)'s facts against \(A\)'s fact store, exactly as a write
would: equivalent facts merge their occurrences, and distinct ones are added.
MemForest then reconciles the scope assignments and cluster states of the two
forests. A matched pair of MemTrees is combined by merging only the touched
subtrees under the fan-out bound, which dirties only the touched paths, while
unmatched trees are copied with their views, embeddings, and index rows
intact. Merging therefore skips extraction, which takes 67\% of a build
(Section~\ref{sec:eval-write-path}), and refreshes only the views whose
evidence changed; external memory is imported as another forest through the
same path. \emph{Re-materialization} regenerates selected
views and indexes from facts, placements, and tree structure when summary or
index settings change, without replaying \(\mathcal{D}_T\).

\noindent\textit{\textbf{Refresh boundaries and visibility.}}
The released user-facing wrapper holds a per-user reentrant lock across
write and query operations. Within a refresh, independent summary requests
run concurrently across nodes and trees. The evaluated batch builder invokes
extraction, canonicalization, structural edits, refresh, and indexing as
separate phases. Its fresh-view completion point is after refresh and index
updates with no unresolved dirty nodes. A failed summary call keeps the
previous view and leaves the node dirty for a later flush; a flush return
alone therefore does not certify freshness. The continuous-update study uses
explicit refresh-and-persist boundaries and queries only at shared completed
checkpoints (Section~\ref{sec:eval-maintenance}).

\section{Evaluation}
\label{sec:evaluation}
\label{sec:ablation}

Our evaluation answers four questions.
\begin{itemize}
    \item \textbf{Q1} (Section~\ref{sec:eval-overall}): Does MemForest build
    memory faster than existing systems without losing answer accuracy, and at
    what cost?
    \item \textbf{Q2} (Section~\ref{sec:eval-write-path}): Where does the
    write speedup come from, and what do decoupled extraction (C1) and
    coalesced, level-parallel refresh (C2) each contribute?
    \item \textbf{Q3} (Section~\ref{sec:eval-query-path}): What do views cost
    and return, and does detail stay reachable under compact views (C3)?
    \item \textbf{Q4} (Sections~\ref{sec:eval-sensitivity}
    and~\ref{sec:eval-migration}): How do the cell size and the fan-out affect
    MemForest, and can built memories be merged without re-ingestion?
\end{itemize}

\subsection{Experimental Setup}
\label{sec:eval-setup}

\noindent\textit{\textbf{Benchmarks.}}
We use two benchmarks. \textsc{LongMemEval-S}~\cite{longmemeval} has 500
questions, each with its own memory instance. \textsc{LoCoMo}~\cite{locomo}
has 10 long multi-session conversations with 1,986 questions. Following the
evaluation protocol of the baseline systems~\cite{mem0,evermemos}, the main
tables exclude its 446 adversarial questions (category 5) and report
categories 1--4.

\noindent\textit{\textbf{Baselines.}}
We compare against six released systems: EverMemOS~\cite{evermemos},
LightMem~\cite{lightmem}, MemoryOS~\cite{memoryos},
MemPalace~\cite{mempalace}, Mem0~\cite{mem0}, and Zep. Zep is reproduced with
Graphiti, Neo4j, and local model services following Zhou et
al.~\cite{zhou2026we}, and we call it Zep Local. Every baseline runs in its
released default configuration, with its native construction, retrieval, and
answer interface and a final budget of ten evidence units. Under these
defaults, EverMemOS answers through agentic search, which iterates retrieval
for three rounds before answering, whereas all other systems retrieve once and
then answer. MemForest also retrieves once, using forest recall followed by
embedding-guided browsing (Section~\ref{sec:query}). Build rates are compared
among the systems whose views become visible when a write returns, which
excludes LightMem and MemPalace.

\noindent\textit{\textbf{Models and hardware.}}
We use three LLM backbones, Qwen3-4B-Instruct-2507,
Qwen3-30B-A3B-Instruct-2507, and Gemma-4-12B-IT, with Qwen3-Embedding-0.6B
\cite{yang2025qwen3,zhang2025qwen3} for all embeddings. Each model is served
on a dedicated NVIDIA H100 GPU with vLLM and
FlashAttention~\cite{dao2023flashattention2}.

\noindent\textit{\textbf{MemForest defaults.}}
MemForest uses cells of \(b=2\) turns, fan-outs \(f\) of 3, 8, and 10 for
session, entity, and scene trees, and view targets of \(W=200\) words, or 250
in entity trees. Queries use \(K=10\), \(M=30\), and \(B=10\). LLM
concurrency is limited to 128, the serving budget of one H100 GPU.
Section~\ref{sec:eval-sensitivity} justifies \(b\) and \(f\).

\noindent\textit{\textbf{Metrics.}}
We report pass@1 accuracy, write-path wall-clock time, and LLM token and call
overhead. Answers are judged by \texttt{deepseek-v4-flash}
\cite{xu2026deepseek} at temperature zero with the public benchmark prompts.
Controlled studies over shared facts use Qwen3-30B grading.
\fullonly{Appendix~\ref{app:revision-protocols} gives prompts and budgets.}

\subsection{Overall Performance}
\label{sec:eval-overall}
\label{sec:main-longmemeval}
\label{sec:main-locomo}

\begin{table}[t]
\centering
\caption{LongMemEval-S pass@1 by question type.}
\label{tab:main-longmemeval}
\scriptsize
\setlength{\tabcolsep}{2.2pt}
\renewcommand{\arraystretch}{0.97}
\resizebox{\columnwidth}{!}{%
\begin{tabular}{lrrrrrrr}
\toprule
Method & Overall & SS-U & SS-P & SS-A & K-Upd. & Multi & Temp. \\
\midrule
\multicolumn{8}{c}{\textbf{Qwen3-4B-Instruct-2507}} \\
\midrule
\rowcolor{gray!15} \textbf{MemForest} & \textbf{72.60} & \textbf{95.71} & 70.00 & 83.93 & 70.51 & 59.40 & \textbf{70.68} \\
EverMemOS & 61.20 & 84.29 & 66.67 & 69.64 & 64.10 & 55.64 & 48.12 \\
Zep Local & 69.60 & 94.29 & 53.33 & \textbf{98.21} & 75.64 & 59.40 & 54.89 \\
LightMem & 68.80 & 94.29 & \textbf{76.67} & 35.71 & \textbf{79.49} & \textbf{65.41} & 64.66 \\
MemoryOS & 64.20 & 87.14 & 46.67 & 89.29 & 50.00 & 57.14 & 60.90 \\
MemPalace & 66.80 & 91.43 & 36.67 & 94.64 & 65.38 & 63.91 & 52.63 \\
Mem0 & 44.80 & 82.86 & 33.33 & 26.79 & 43.59 & 42.11 & 38.35 \\
\midrule
\multicolumn{8}{c}{\textbf{Qwen3-30B-A3B-Instruct-2507}} \\
\midrule
\rowcolor{gray!15} \textbf{MemForest} & \textbf{81.80} & \textbf{97.14} & 83.33 & 83.93 & 75.64 & \textbf{76.69} & \textbf{81.20} \\
EverMemOS & 67.00 & 91.43 & 53.33 & 75.00 & 79.49 & 62.41 & 51.13 \\
Zep Local & 71.00 & 95.71 & 63.33 & \textbf{96.43} & \textbf{82.05} & 56.39 & 57.14 \\
LightMem & 70.20 & 95.71 & \textbf{86.67} & 37.50 & 79.49 & 64.66 & 66.92 \\
MemoryOS & 63.80 & 87.14 & 60.00 & 89.29 & 62.82 & 52.63 & 53.38 \\
MemPalace & 59.80 & 81.43 & 50.00 & 91.07 & 62.82 & 52.63 & 42.86 \\
Mem0 & 50.20 & 91.43 & 46.67 & 26.79 & 60.26 & 42.86 & 40.60 \\
\midrule
\multicolumn{8}{c}{\textbf{Gemma-4-12B-IT}} \\
\midrule
\rowcolor{gray!15} \textbf{MemForest} & 77.80 & 95.71 & 63.33 & 46.43 & 82.05 & 74.44 & \textbf{85.71} \\
EverMemOS & \textbf{78.00} & 95.71 & 83.33 & 76.79 & 84.62 & 72.93 & 69.17 \\
Zep Local & 75.60 & 94.29 & 66.67 & \textbf{96.43} & \textbf{89.74} & 63.16 & 63.16 \\
LightMem & 77.00 & \textbf{97.14} & \textbf{86.67} & 37.50 & 80.77 & \textbf{75.19} & 80.45 \\
MemoryOS & 66.00 & 95.71 & 70.00 & 94.64 & 66.67 & 47.37 & 55.64 \\
MemPalace & 72.80 & 88.57 & 50.00 & 94.64 & 79.49 & 60.90 & 68.42 \\
Mem0 & 59.20 & 91.43 & 50.00 & 23.21 & 74.36 & 54.89 & 54.89 \\
\bottomrule
\end{tabular}}
\par\vspace{4pt}
\parbox{\columnwidth}{\scriptsize\raggedright SS-U, SS-P, SS-A: single-session user, preference,
and assistant. K-Upd.: knowledge update. Multi: multi-session. Temp.: temporal
reasoning.}
\end{table}

\begin{table}[t]
\centering
\caption{LoCoMo pass@1 on categories 1--4.}
\label{tab:main-locomo}
\scriptsize
\setlength{\tabcolsep}{3.0pt}
\renewcommand{\arraystretch}{0.97}
\resizebox{\columnwidth}{!}{%
\begin{tabular}{lrrrrr}
\toprule
Method & Overall & Single-hop & Multi-hop & Open-ended & Temporal \\
\midrule
\multicolumn{6}{c}{\textbf{Qwen3-4B-Instruct-2507}} \\
\midrule
\rowcolor{gray!15} \textbf{MemForest} & 78.12 & 80.86 & \textbf{81.21} & \textbf{53.12} & 75.70 \\
EverMemOS & \textbf{79.22} & \textbf{83.71} & 79.79 & 47.92 & \textbf{76.32} \\
Zep Local & 68.70 & 78.36 & 62.77 & 34.38 & 58.88 \\
LightMem & 66.17 & 69.68 & 68.79 & 40.62 & 62.31 \\
MemoryOS & 65.45 & 71.70 & 71.99 & 39.58 & 51.09 \\
MemPalace & 51.56 & 55.77 & 68.79 & 29.17 & 32.09 \\
Mem0 & 47.86 & 45.66 & 52.84 & 39.58 & 51.71 \\
\midrule
\multicolumn{6}{c}{\textbf{Qwen3-30B-A3B-Instruct-2507}} \\
\midrule
\rowcolor{gray!15} \textbf{MemForest} & 84.09 & 84.66 & \textbf{86.88} & \textbf{67.71} & 85.05 \\
EverMemOS & \textbf{84.22} & \textbf{87.63} & 84.04 & 51.04 & \textbf{85.36} \\
Zep Local & 76.75 & 82.52 & 76.95 & 56.25 & 67.60 \\
LightMem & 60.52 & 62.07 & 62.06 & 44.79 & 59.81 \\
MemoryOS & 68.31 & 75.98 & 72.34 & 50.00 & 50.16 \\
MemPalace & 55.84 & 59.81 & 72.70 & 33.33 & 37.38 \\
Mem0 & 59.22 & 58.86 & 61.35 & 58.33 & 58.57 \\
\midrule
\multicolumn{6}{c}{\textbf{Gemma-4-12B-IT}} \\
\midrule
\rowcolor{gray!15} \textbf{MemForest} & 82.66 & 80.98 & 89.36 & 57.29 & 88.79 \\
EverMemOS & \textbf{88.57} & \textbf{90.84} & \textbf{89.72} & \textbf{63.54} & \textbf{89.10} \\
Zep Local & 69.55 & 74.91 & 68.79 & 26.04 & 69.16 \\
LightMem & 78.90 & 81.33 & 79.43 & 50.00 & 80.69 \\
MemoryOS & 78.51 & 80.98 & 83.33 & 57.29 & 74.14 \\
MemPalace & 55.71 & 58.98 & 77.30 & 28.12 & 36.45 \\
Mem0 & 48.38 & 44.71 & 57.09 & 39.58 & 52.96 \\
\bottomrule
\end{tabular}}
\end{table}

\noindent\textit{\textbf{Answer accuracy.}}
Tables~\ref{tab:main-longmemeval} and~\ref{tab:main-locomo} report pass@1
accuracy on the two benchmarks under the three backbones. On LongMemEval-S,
MemForest is the most accurate system under both Qwen backbones. It reaches
81.8\% with Qwen3-30B and 72.6\% with Qwen3-4B, which is 10.8 and 3.0 points
above Zep Local, the strongest baseline in both cases. It also leads the
temporal-reasoning category under all three backbones, by 14.3 points with
Qwen3-30B, which is consistent with its time-ordered trees and the event times
it renders with every fact. On LoCoMo, MemForest reaches 84.1\% with Qwen3-30B
and is statistically indistinguishable from EverMemOS, the strongest baseline:
the difference is \(-0.13\) points, with a 95\% confidence interval of
\([-2.2,+2.1]\). It also leads the multi-hop and open-ended categories under
both Qwen backbones, by 16.7 points over EverMemOS on the open-ended questions
with Qwen3-30B.
\fullonly{LoCoMo's 446 adversarial questions (category 5) are scored
separately with a strict answerability judge, because the public prompts assume
a non-empty answer. On these questions MemForest reaches 35.9\% with Qwen3-30B
against 19.7\% for EverMemOS, while MemoryOS scores highest at 44.8\%.}

EverMemOS is the only baseline that MemForest does not lead in every setting.
Its agentic search spends three retrieval rounds on each question, whereas
MemForest retrieves once. Even so, MemForest is ahead of EverMemOS or within
1.1 points of it in every setting except LoCoMo under Gemma, where it trails by
5.9 points. Among the other baselines, which also retrieve once, MemForest is
the most accurate on both benchmarks under all three backbones, so a single
retrieval pass over its memory is competitive with agentic search.

MemForest's weakest result is the single-session-assistant category under
Gemma, at 46.4\% against 83.9\% under both Qwen backbones. These questions ask
about what the assistant said. Zep Local, MemoryOS, and MemPalace keep the raw
dialogue retrievable and score 89--98\% under every backbone, so the answer
model is not the cause. MemForest answers from extracted facts, and the drop
comes from extraction, where Gemma keeps less assistant-side detail than Qwen.
Only 11.8\% of the facts Gemma extracts originate from assistant turns,
against 69.7\% with Qwen3-30B, although both extract about 1.5 facts per user
turn.

\begin{figure}[t]
    \centering
    \includegraphics[width=\columnwidth]{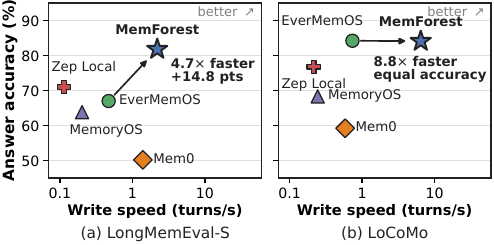}
    \caption{Write speed and answer accuracy with Qwen3-30B.}
    \label{fig:speed-accuracy}
\end{figure}

\noindent\textit{\textbf{Write speed.}}
Figure~\ref{fig:speed-accuracy} places each system by how fast it writes and
how accurately it answers with Qwen3-30B. On LongMemEval-S, the baselines
trade one for the other: Zep Local is the most accurate and the slowest, while
Mem0 is the fastest and the least accurate. MemForest escapes this trade-off.
It writes \(1.6\times\) faster than Mem0 while answering 10.8 points above Zep
Local. Compared with EverMemOS, the baseline closest to it in accuracy,
MemForest builds memory \(4.7\times\) faster on LongMemEval-S while answering
14.8 points higher, and \(8.8\times\) faster on LoCoMo at equal accuracy,
measured on one 680-turn conversation. Because facts are stored apart from
views, a faster write path need not cost answer quality.

\begin{table}[t]
\centering
\caption{Mean billable token use and direct API cost on the
DeepSeek-V4-Flash probe, and query latency with Qwen3-30B. Tokens are in
thousands; higher cost efficiency is better.}
\label{tab:cache-token-probe-main}
\setlength{\tabcolsep}{2.8pt}
\resizebox{\columnwidth}{!}{
\begin{tabular}{lrrrrrrr}
\toprule
Method & Cached in & Miss in & Output & Total & Cost (\textcent) & Cost eff. & Latency (s) \\
\midrule
\rowcolor{gray!15}
\textbf{MemForest} & 15.45 & 7.51 & 3.64 & 26.60 & 0.212 & 5.28$\times$ & 2.2 \\
EverMemOS & 3.37 & 24.57 & 3.45 & 31.40 & 0.442 & 2.53$\times$ & 8.5 \\
Mem0 & 13.65 & 4.44 & 1.27 & 19.37 & 0.102 & 11.00$\times$ & 0.2 \\
MemoryOS & 1.88 & 8.84 & 2.44 & 13.15 & 0.193 & 5.81$\times$ & 2.3 \\
Zep Local & 47.66 & 66.77 & 6.06 & 120.48 & 1.118 & 1.00$\times$ & 2.8 \\
\bottomrule
\end{tabular}
}
\end{table}

\noindent\textit{\textbf{Cost and query latency.}}
Table~\ref{tab:cache-token-probe-main} reports the billable tokens, API cost,
and query latency of each system, where tokens and cost come from a
cache-warmed probe that covers ingestion, retrieval, and answering. MemForest
costs 0.21 cents, which is 52\% less than EverMemOS and \(5.3\times\) less
than Zep Local, because its requests are short and its stable refresh prompts
make 58\% of its tokens cached input. Mem0 and MemoryOS cost less, but they
are the two least accurate of these five systems and trail MemForest by 31.6
and 18.0 points on LongMemEval-S with Qwen3-30B. At query time, MemForest
answers a question in 2.2 seconds, on par with MemoryOS and Zep Local and
\(3.9\times\) faster than EverMemOS; only Mem0 is faster. Retrieval takes 0.5
of these seconds and does not slow down as the memory grows: when the index
grows from 1,428 to 10,781 facts, the median index lookup stays below
40\,ms.

\begin{table}[t]
\centering
\scriptsize
\caption{Evidence-missing errors per system. \(N\) counts wrong
answers whose answer context lacks the supporting evidence.}
\label{tab:error-attribution}
\setlength{\tabcolsep}{2.6pt}
\begin{tabular}{lrrr}
\toprule
System & $N$ & \shortstack{Not found in\\audited store units} & \shortstack{Stored, not in\\answer context} \\
\midrule
\rowcolor{gray!15}
\textbf{MemForest} & 11 & 6 (54.5\%) & 5 (45.5\%) \\
EverMemOS & 21 & 13 (61.9\%) & 8 (38.1\%) \\
Zep Local & 25 & 0 (0.0\%) & 25 (100.0\%) \\
MemPalace & 31 & 0 (0.0\%) & 31 (100.0\%) \\
MemoryOS & 56 & 5 (8.9\%) & 51 (91.1\%) \\
LightMem & 38 & 7 (18.4\%) & 31 (81.6\%) \\
Mem0 & 83 & 49 (59.0\%) & 34 (41.0\%) \\
\bottomrule
\end{tabular}
\end{table}

\noindent\textit{\textbf{Where the remaining errors come from.}}
\label{sec:eval-error-attribution}
Table~\ref{tab:error-attribution} attributes the wrong answers of each system
to the stage that loses their evidence. On a stratified sample of 100
LongMemEval-S and 100 LoCoMo questions from the Qwen3-30B runs, we take every
wrong answer whose answer context lacks the supporting evidence and check
whether that evidence is still in the store. MemForest has the fewest such
errors, 11, against 21 for EverMemOS and 83 for Mem0. Mem0 mostly fails to
preserve evidence: for 59\% of its errors, the evidence is not in the store.
Zep Local, MemPalace, MemoryOS, and LightMem mostly fail to deliver it: for
more than 80\% of their errors, the evidence is stored but never reaches the
answer context. Accuracy thus needs both stored evidence and an access path
that delivers it, and MemForest provides both by keeping every extracted fact
and reaching it through fact-to-tree recall and browsing.
\fullonly{Appendix~\ref{app:error-attribution-audit} gives the audit protocol and
trace provenance.}

\subsection{Write-Path Analysis}
\label{sec:eval-write-path}
\label{sec:eval-maintenance}
\label{sec:ablation-maintenance}
\label{sec:eval-ablation}

To answer Q2, we break build time down by stage and then isolate decoupled
extraction (C1), batch coalescing, and level-parallel refresh (C2).

\begin{figure}[t]
    \centering
    \includegraphics[width=\columnwidth]{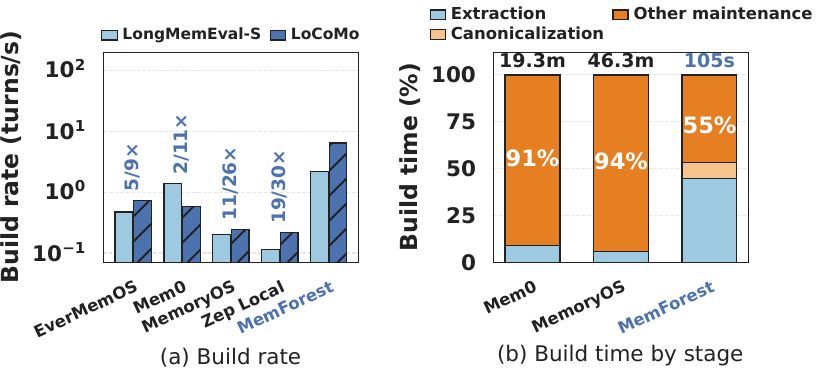}
    \caption{Write path with Qwen3-30B: build rate and build time by stage.}
    \label{fig:write}
\end{figure}

\noindent\textit{\textbf{Where build time goes.}}
Figure~\ref{fig:write}a reports the build rate of MemForest and of the four
baselines with immediately visible views on both benchmarks. MemForest is the
fastest in every case, from \(1.6\times\) faster than Mem0 on LongMemEval-S to
\(30.0\times\) faster than Zep Local on LoCoMo. The gap to Mem0 widens to
\(11\times\) on LoCoMo partly because its benchmark adapter issues one
user-scoped write per named speaker.
Figure~\ref{fig:write}b breaks the build time of one LoCoMo conversation down
by stage. Mem0 and MemoryOS need 19.3 and 46.3 minutes and spend 91\% and
94\% of their LLM time on maintenance, the dependency chains measured in
Section~\ref{subsec:motivation_study}.
MemForest builds the same conversation in 105 seconds, with 47 seconds of
extraction and 58 seconds of maintenance. The split depends on the workload:
on twelve LongMemEval-S instances, extraction takes 67\% of the 229-second
build and view refresh only 19\%. In both cases, the build no longer waits on
a long chain of dependent calls. Mem0's longest chains contain 40--43
dependent calls, whereas MemForest's extraction calls are mutually independent
and its refresh depth is bounded by tree height, at most 4 in our builds.

\noindent\textit{\textbf{Decoupled extraction (C1).}}
To isolate extraction parallelism, we replay the extraction requests of three
LongMemEval-S write traces serially, at calibrated throughput, and compare
with their parallel schedule. The requests and their outputs are identical, so
only the schedule changes. The serial replay takes
54.6 minutes and the parallel schedule 122 seconds, a \(26.8\times\)
reduction, which is possible because every extraction call reads only its own
cell and never the existing memory.

\begin{figure}[t]
    \centering
    \includegraphics[width=\columnwidth]{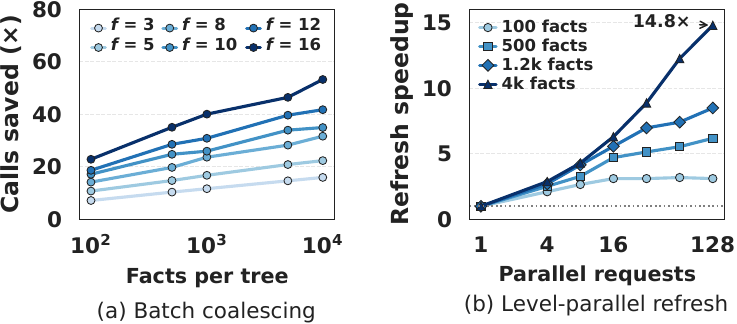}
    \caption{Batch coalescing and level-parallel refresh in MemTrees.}
    \label{fig:maint}
\end{figure}

\noindent\textit{\textbf{Batch coalescing (C2).}}
Figure~\ref{fig:maint}a plots the summary calls that coalescing saves over
refreshing once per insert, for trees of \(10^2\) to \(10^4\) facts and
fan-outs from 3 to 16. The saving ranges from \(7\times\) to \(53\times\) and
grows with both tree size and fan-out, because more inserts then fall under
the same ancestors.
The refresh boundary determines how much can be coalesced. On
continuous-update streams of 128 events over each of three histories,
refreshing every 32 events instead of every 8 cuts refresh tokens by 70\%.
Both schedules give the same answers to the 192 checkpoint questions asked
every 32 events, so a larger batch gives up only the intermediate view
versions between checkpoints.

\begin{figure}[t]
    \centering
    \includegraphics[width=\columnwidth]{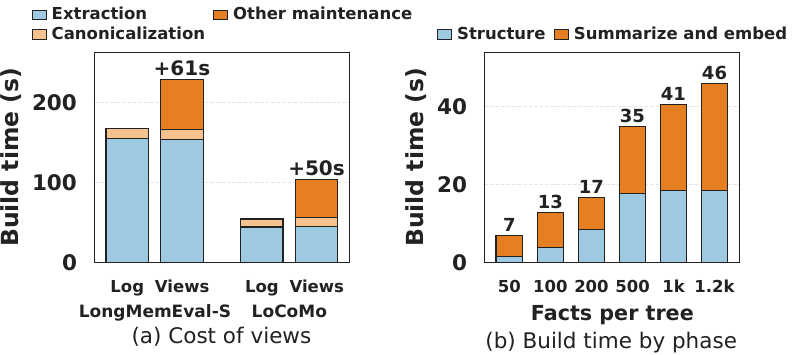}
    \caption{View maintenance and construction cost.}
    \label{fig:writecost}
\end{figure}

\noindent\textit{\textbf{Level-parallel refresh (C2).}}
Recomputing a view from its current children lets all dirty nodes of a level,
across all trees, refresh together. Figure~\ref{fig:maint}b shows the speedup
of this level-parallel refresh over serial refresh as the number of parallel
requests grows from 1 to 128. At \(P=128\), the speedup reaches \(8.5\times\)
for 1,200-fact trees and \(14.8\times\) for 4,000-fact trees. The gain thus
grows with scope size, since a larger tree has more dirty nodes per level. On
the three LongMemEval-S write traces, parallel refresh takes 28 seconds
against 9.9 minutes for a calibrated serial refresh, a \(21.2\times\)
reduction. Figure~\ref{fig:writecost}b shows the resulting scaling: growing
one tree from 50 to 1,200 facts, a \(24\times\) increase, raises build time
by only \(6.7\times\).

\subsection{Query-Path Analysis}
\label{sec:eval-query-path}
\label{sec:eval-same-pipeline}
\label{tab:same-pipeline}
\label{sec:eval-navigation}
\label{sec:short-root-pilot}
\label{sec:ablation-retrieval}
\label{sec:ablation-scaling}

To answer Q3, we measure what views cost for point lookups, what they return
for range aggregation, and whether detail stays reachable under short views.

\noindent\textit{\textbf{Cost of views.}}
In Figure~\ref{fig:writecost}a, we compare MemForest against
\emph{MemForest-Log}, a variant without views that stores the same canonical
facts in time order and retrieves them with tuned semantic, lexical, and
temporal scoring, on \RthreeUnitsLME{} LongMemEval-S instances and
\RthreeUnitsLoCoMo{} LoCoMo conversations. Views add 36\% to the build time on
LongMemEval-S and 91\% on LoCoMo, together with 46\% more build tokens.
Point-query accuracy is unaffected. At evidence budgets from 10 facts to
2,048 tokens, the overall accuracy of the two variants stays within 1.1
points, and no difference is significant under paired McNemar tests
(\(p\ge0.18\)). Views thus neither help nor hurt point lookups, which a few
facts answer; their return comes from aggregation, which we measure next.

\begin{figure}[t]
    \centering
    \includegraphics[width=\columnwidth]{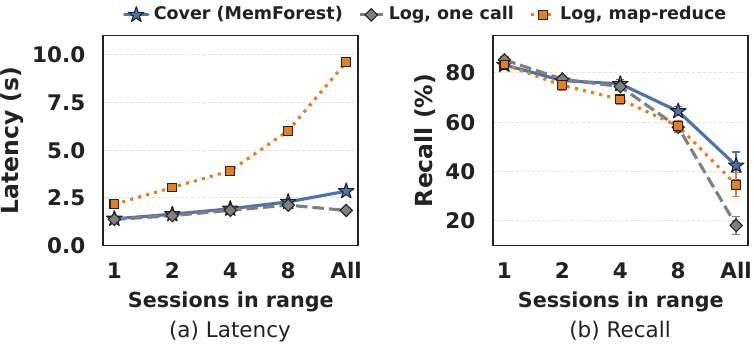}
    \caption{Time-range summaries on LoCoMo entity trees.}
    \label{fig:range}
\end{figure}

\noindent\textit{\textbf{Range summaries.}}
Neither benchmark issues aggregate requests, so we construct 499 requests of
the form ``what happened to \(X\) between \(t_1\) and \(t_2\)'' over the
per-speaker entity trees of the ten LoCoMo conversations, with ranges from
one session to the full history. In Figure~\ref{fig:range}, we compare three
ways to answer them. \emph{Cover} merges the cover views of
Proposition~\ref{prop:cover} in one LLM call. The two log arms summarize the
in-range facts at query time, in one call or by map-reduce, under the same
model, prompt, and word cap; recall is measured as in
Figure~\ref{fig:fanout}a.
Cover answers within 3 seconds at every range length. Map-reduce is up to
\(3.2\times\) slower and reads up to \(30\times\) more input tokens, and from
four sessions on it also recalls fewer facts. A single call over the log is as
fast as Cover and matches its recall up to four sessions, but its recall
collapses on long ranges. It recalls 0.58 of the facts against 0.64 for Cover
at eight sessions, and 0.18 against 0.42 on the full history. The cover itself
stays small, with at most 25 views, within the bound of
Proposition~\ref{prop:cover}. Reusing materialized views therefore pays off
as ranges grow; for one or two sessions the paths cost about the same.

\begin{figure}[t]
    \centering
    \includegraphics[width=\columnwidth]{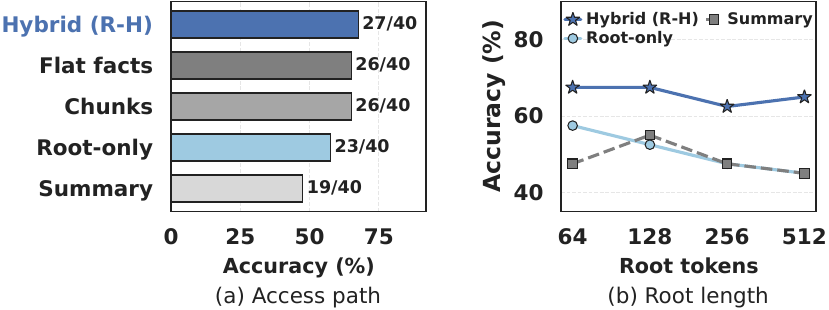}
    \caption{MemTrees as a navigation index.}
    \label{fig:nav}
\end{figure}

\noindent\textit{\textbf{Detail under compact roots.}}
Compact views are useful only if the detail they omit stays reachable. On
three LoCoMo histories, we freeze the
facts, the tree topology, and the non-root summaries, regenerate all 313 roots
at targets from 512 down to 64 tokens, and answer the same 40 questions.
Figure~\ref{fig:nav}a compares the access paths at the 64-token target. Root
summaries alone answer 19 of the 40 questions and descent from the recalled
roots 23. The hybrid path R-H adds fact-to-tree recall and, unlike the native
interface of Section~\ref{sec:query}, direct top-30 fact candidates; it
answers 27, on par with the flat-fact and chunk-index controls at 26.
\fullonly{Appendix~\ref{app:short-root-pilot} gives the 2,048-token packing protocol.}
Figure~\ref{fig:nav}b varies the root length. Hybrid access is insensitive to
it: it stays at 25--27 correct answers across the four targets, and shrinking
the target from 512 to 64 tokens leaves 36 of the 40 final answer contexts
unchanged, whereas root-only descent varies between 18 and 23. Roots can therefore be kept
short, because facts and fact-to-tree links, rather than the root text, carry
the detail.

\subsection{Parameter Sensitivity}
\label{sec:eval-sensitivity}

For the first part of Q4, we justify the defaults of the cell size \(b\) and
the fan-out \(f\).

\begin{figure}[t]
    \centering
    \includegraphics[width=\columnwidth]{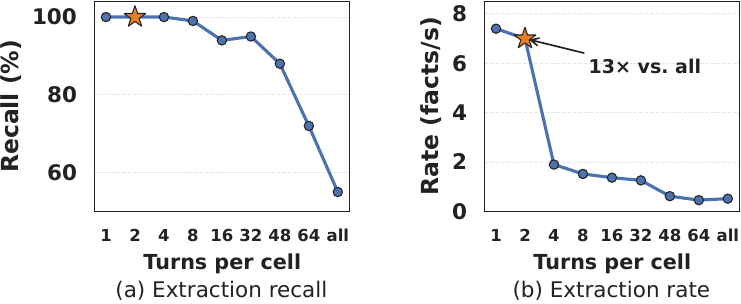}
    \caption{Extraction recall and throughput versus extraction-cell size.}
    \label{fig:chunks}
\end{figure}

\noindent\textit{\textbf{Extraction-cell size.}}
\label{app:chunk-sweep}
Figure~\ref{fig:chunks} varies the number of turns per extraction cell, from
one turn to a whole session, on long-session stress cases. It reports the
extraction rate and the extraction recall, the share of questions whose
extracted facts still contain the answer-bearing span of the gold turns.
Two-turn cells cover every gold span at 7.0 facts per second. Recall falls
beyond 8 turns per cell, and extracting a whole session in a single pass
covers only 55\% of the gold spans, is \(13\times\) slower, and spends
\(3.3\times\) the tokens per fact. One-turn cells match the recall of two-turn
cells but spend 13\% more tokens per fact, so we use \(b=2\).

\begin{figure}[t]
    \centering
    \includegraphics[width=\columnwidth]{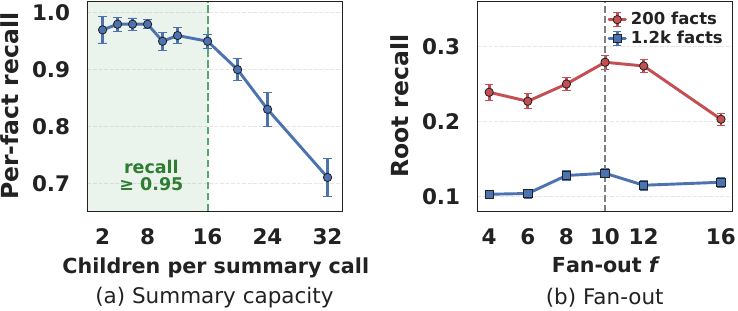}
    \caption{Summary capacity and fan-out.}
    \label{fig:fanout}
\end{figure}

\noindent\textit{\textbf{Fan-out.}}
The fan-out trades summary capacity against tree depth. In
Figure~\ref{fig:fanout}a, over 30 groups of facts, one summary call keeps
per-fact recall at or above 0.95 for up to 16 children and degrades beyond. A
small fan-out is not better either, because deeper trees lose detail across
levels. Figure~\ref{fig:fanout}b shows the combined effect on root recall over
56 histories, which peaks at \(f=10\) for both 200-fact and 1,200-fact trees.
We therefore set \(f=10\) for scene trees, \(f=8\) for entity trees, and
\(f=3\) for session trees, whose leaves are cells several facts long.

\subsection{Efficient Migration}
\label{sec:eval-migration}

The second part of Q4 concerns reuse after a memory is built, for example
when it is moved to a new deployment or synchronized across devices. We merge
one to eight independently built LongMemEval-S memories into one state, either
by replaying every source history through the write path or by combining the
materialized forests through the merge path of Section~\ref{sec:lifecycle}.
Merging is faster at every scale, by more than \(2\times\) from three
instances on and by up to \(2.70\times\), because it never re-extracts
processed dialogue and copies unmatched trees with their views intact. The
speedup does not come from discarding state: at eight instances, the merged
memory holds 10,801 facts, against 10,705 for the replayed one. No evaluated
baseline can merge built memories without re-ingesting the source histories.

\section{Related Work}
\label{sec:related}

\noindent\textit{\textbf{View maintenance and holistic aggregates.}}
Incremental view maintenance propagates base-table deltas to materialized
views \cite{gupta1995maintenance}. Its delta rules rely on algebraic
structure: distributive and algebraic aggregates can be combined from partial
results, whereas a holistic aggregate generally needs recomputation from its
underlying groups \cite{gray1996datacube,gray1997data}. LLM summaries similarly require a policy for keeping and recomputing
information: their generated text supplies no prescribed exact delta rule.
Temporal aggregation trees come closest in structure: the aggregation tree
computes aggregates over constant intervals \cite{kline1995temporal}, and the
SB-tree maintains temporal aggregates incrementally by storing partial values
on a balanced tree over time \cite{yang2001sbtree}. Both rely on
distributive or algebraic aggregates, whose partial values can be combined
and updated by deltas. Write-optimized and temporal indexes avoid rewriting
hot state and keep historical versions
\cite{lsmtree,elmasri1990time,becker1996asymptotically}.
MemForest carries these ideas to holistic LLM views: it materializes
interval views in a time-ordered hierarchy, recomputes an affected view from
its current children instead of applying a delta, reuses unaffected children,
answers range summaries from a cover of interval views, and keeps base evidence
immutable under refresh.

\noindent\textit{\textbf{Database support for LLM applications.}}
AquaPipe and Cache-Craft optimize retrieval--generation execution, and
GaussDB and HAKES improve vector search
\cite{10.1145/3709661,10.1145/3725273,sun2025gaussdb,hu2025hakes}; these
engines can serve MemForest's retrieval, while MemForest maintains the memory
and its derived indexes under continuous writes.

\noindent\textit{\textbf{Agent memory systems.}}
Mem0 extracts facts and reconciles them with retrieved records
\cite{mem0}; MemoryOS stages short-, mid-, and long-term memory with segment
and profile updates \cite{memoryos}; EverMemOS consolidates episodes from a
streaming segmenter \cite{evermemos}; and Zep resolves new nodes and edges
against a temporal knowledge graph \cite{zep,zhou2026we}. SeCom, A-Mem, RMM,
and context-dependent frameworks study memory units, linking, and reflection
\cite{pansecom,amem,RMM,gao2025efficient}. Table~\ref{tab:design_space} contrasts their write dependencies. LightMem is closest in decoupling extraction:
it buffers updates and consolidates them offline in parallel
\cite{lightmem}, but its views become visible only after consolidation,
whereas MemForest recomputes affected views at an explicit write-batch
boundary.
MemPalace keeps near-verbatim history and defers abstraction to query time
\cite{mempalace}.

\noindent\textit{\textbf{Hierarchical memory.}}
RAPTOR recursively summarizes document clusters into a static tree
\cite{sarthi2024raptor}. Dynamic Tree Memory inserts new content by
aggregating it into existing ancestor representations, i.e., by revising each
ancestor on the path, and permits parallel aggregation along that path
\cite{rezazadehisolated}. H-Mem, HiMem, SHIMI, Collaborative Memory, and
transactive memory likewise organize memory hierarchically or across agents
\cite{hmem,himem,helmi2025decentralizing,rezazadeh2025collaborative,
jackson2008transactive,Adapt,hu2025hiagent}. A MemForest refresh
instead reads the current children, not the previous view, and the same
placements both scope maintenance and route fact hits back to trees; GraphRAG \cite{edge2024local} organizes evidence at a
different granularity.

\section{Conclusion}
\label{sec:conclusion}

We presented MemForest, an agent memory system that keeps facts as durable
evidence and derives every semantic view from them. Facts are extracted
independently of existing memory, and each view is recomputed from its
current children, so refresh depth follows tree height rather than history
length. Fact-to-tree links keep omitted detail reachable, and interval views
serve time-range aggregation. With Qwen3-30B, MemForest builds memory
4.7--8.8\(\times\) faster than EverMemOS at equal or higher accuracy. These
results suggest that LLM-generated state is better derived from durable
evidence than revised in place, a principle we plan to extend to memory
shared across agents.

\section*{AI-Generated Content Acknowledgment}
The authors used Claude Opus 5.5 (Anthropic) through Claude Code, OpenAI
Codex, and ChatGPT. These systems edited text throughout the paper and drafted
some passages. They also wrote and ran analysis, plotting, and benchmark
scripts for the results in Sections~\ref{sec:background}
and~\ref{sec:evaluation}, including the charts in
Figures~\ref{fig:motivation} and~\ref{fig:speed-accuracy}--\ref{fig:fanout}.
The authors directed this work, reviewed, tested, and validated all
AI-assisted scripts and text, and are responsible for the experimental
results, scientific claims, and originality of this work.

\clearpage
\bibliographystyle{ACM-Reference-Format}
\bibliography{reference}

\clearpage
\appendix
\counterwithin{figure}{section}
\counterwithin{table}{section}
\renewcommand{\thefigure}{\thesection\arabic{figure}}
\renewcommand{\thetable}{\thesection\arabic{table}}
\renewcommand{\theHfigure}{appendix.\thesection.\arabic{figure}}
\renewcommand{\theHtable}{appendix.\thesection.\arabic{table}}
\section{Write-Path and Cost Measurement Details}
\label{app:detailed-results}

\begin{table}[!htbp]
\centering
\scriptsize
\caption{Share of native build work spent on extraction and on
state-dependent maintenance (Qwen3-30B, one H100).}
\label{tab:stage-share}
\setlength{\tabcolsep}{3.4pt}
\begin{tabular}{lrrr}
\toprule
Method & Extraction & Maintenance & Maint.\ output tokens \\
\midrule
Mem0 & 8.9\% & \MemZeroMaintTime & \MemZeroMaintOut \\
MemoryOS & 6.0\% & \MemOSMaintTime & \MemOSMaintOut \\
\rowcolor{gray!15}
\textbf{MemForest} & \MFExtractTime & \MFMaintTime & \MFMaintOut \\
\bottomrule
\end{tabular}
\end{table}

In Table~\ref{tab:stage-share}, the shares of Mem0 and MemoryOS are summed
serial LLM call latency over three LoCoMo histories. The shares of MemForest are
stage wall time, with canonicalization, routing, refresh, and indexing counted
as maintenance.

\noindent\textit{\textbf{Build-rate notes.}}
In Figure~\ref{fig:speed-accuracy}, source-turn build rates are comparable only
within a benchmark. Mem0's lower LoCoMo rate follows the evaluated EverMemOS
benchmark adapter: named speakers activate its dual-perspective path, which
awaits two user-scoped writes sequentially (pinned \texttt{online\_base.py},
lines 158--174, commit 539db77). We report the released workflow without
adding a baseline-specific optimization.

\noindent\textit{\textbf{Cost-probe protocol.}}
Table~\ref{tab:cache-token-probe-main} uses five
conversations from the frozen length-stratified LoCoMo subset: three measured
segments and two disjoint cache-warmup segments, each truncated to its first
20 messages. Every method starts from an empty memory and runs its native
ingestion, retrieval, and answer path with DeepSeek-V4-Flash in non-thinking
mode and Qwen3-Embedding-0.6B. A distinct provider \texttt{user\_id} isolates
each method; template validation and the second warmup must show provider
cache hits before measurement. The table reports the unweighted mean of the
three probes; all 15 method--probe cells finish without failed requests. Costs
use API-returned billable token classes and official
pricing~\cite{deepseek-api-pricing}.

\section{Evidence-Missing Error Audit}
\label{app:additional}
\label{app:error-attribution-audit}

Table~\ref{tab:error-attribution} uses a shared stratified sample of 100
questions per benchmark (seed 20261107) and the archived answer-correctness
labels. Among wrong answers judged to lack supporting benchmark evidence
in their recorded context, we distinguish evidence not found in the audited
store units from evidence found in memory but absent from the context.
The former uses BM25 top-250 retrieval followed by embedding reranking;
the judge receives up to 30 units. Full-turn verbatim matches override
negative evidence judgments. Percentages use the sum of these two groups.
The 267 analyzed system--question cases exclude 186 evidence-present wrong
answers and 21 cases with unavailable stores.

Two MemoryOS cases initially lacked a readable mid-term snapshot because
the JSON file ended inside a record. Decoding complete records before the
truncated tail recovered the supporting turns and dates for both questions,
placing both in the stored-but-absent group. LightMem's 21 excluded cases
are all from LongMemEval-S; its available extraction logs do not establish
the complete stored state. Its reported breakdown therefore covers LoCoMo.

Trace provenance is recorded per question: MemForest uses archived
answer labels and sibling-rerun contexts; MemoryOS uses
reconstructed full-page contexts. The artifact records exact and
reconstructed sources for the other methods, the two recovered MemoryOS
records, and all inclusion decisions. These labels locate evidence gaps
in the audited traces; they do not distinguish routing, ranking, browsing,
and context packing within the delivery stage. On correct answers, the
same evidence check flags missing support in 28.6--36.8\% of LightMem,
Mem0, and MemoryOS cases, versus 5.2--11.8\% for the other methods;
the artifact keeps this calibration alongside the breakdown.

\section{Evaluation Protocols}
\label{app:prompts}
\label{app:revision-protocols}

The main tables judge each frozen answer once with
\texttt{deepseek-v4-\allowbreak flash}, temperature zero, and thinking
disabled, using the public Mem0 benchmark prompts~\cite{mem0memorybenchmarks}:
the LongMemEval prompt at
\href{https://github.com/mem0ai/memory-benchmarks/blob/7ba1bd330f6ef6acdc751b6e1f82ac8af0568873/benchmarks/longmemeval/prompts.py}{commit
\texttt{7ba1bd3}} and the tuned LoCoMo prompt at
\href{https://github.com/mem0ai/memory-benchmarks/blob/edcd6f1d42400837b1fcb6997716f1769dc51a37/benchmarks/locomo/prompts.py}{commit
\texttt{edcd6f1}}. LoCoMo category 5 uses a strict answerability prompt,
released verbatim with its hash and rendering code in the
\artifacttree{reproducibility/evaluation}{evaluation package}.

\begingroup
\subsection{Retrieval and Answer Interfaces}

Flat-item systems in the main tables expose a global final
top-\(k=10\) context. We preserve each method's schema-aware answer interface
because the returned objects differ: MemForest returns canonical facts and
tree-derived evidence, EverMemOS returns episodic and recollection contexts, and
other flat systems expose their native memory records. MemForest uses its
native top-\(k=10\) tree-browse setting and fully expands selected tree units;
equal-budget comparisons over shared facts are confined to
Section~\ref{sec:eval-same-pipeline}. Zep Local is not
described as an equal ten-fact comparison: its native Graphiti search uses a
limit of 10 per evidence class and serializes heterogeneous edges, nodes,
episodes, and communities. Table~\ref{tab:zep-native-budget} reports the
resulting object counts and context-token lengths separately.

Following Zhou et al.~\cite{zhou2026we}, the Zep Local row uses the
\artifacttree{reproducibility/baselines/zep_local}{released Graphiti and Neo4j
adapter}: Graphiti v0.24.1, Neo4j 5.26.2, raw-episode ingestion, and self-hosted
generation and Qwen3-Embedding-0.6B endpoints. It is a reproducible local
Graphiti realization, not Zep Cloud. The
\artifactblob{reproducibility/results/zep_local/native_budget_summary.csv}{six-cell
budget summary}, its
\artifactblob{reproducibility/results/zep_local/native_budget_manifest.json}{source
manifest}, and the
\artifactblob{reproducibility/scripts/summarize_zep_native_budget.py}{aggregation
script} preserve the table's result-to-code path.

\begin{table*}[t]
\centering
\caption{Zep Local native retrieval budget. Object columns
report the mean number serialized per question. Context tokens use the exact
serialized context and one fixed Qwen tokenizer; they exclude the answer
instruction and question.}
\label{tab:zep-native-budget}
\setlength{\tabcolsep}{4pt}
\begin{tabular}{llrrrrrr}
\toprule
Backbone & Benchmark & Edges & Nodes & Episodes & Communities &
Context tokens avg. & Context tokens p95 \\
\midrule
Qwen3-4B & LongMemEval-S & 5.00 & 5.00 & 9.99 & 0.00 & 3,269 & 4,542 \\
Qwen3-4B & LoCoMo & 5.00 & 4.97 & 10.00 & 0.00 & 1,360 & 1,626 \\
Qwen3-30B & LongMemEval-S & 5.00 & 5.00 & 9.99 & 0.00 & 3,125 & 4,412 \\
Qwen3-30B & LoCoMo & 5.00 & 5.00 & 10.00 & 0.00 & 1,210 & 1,413 \\
Gemma-4-12B-IT & LongMemEval-S & 5.00 & 5.00 & 9.99 & 0.00 & 3,105 & 4,364 \\
Gemma-4-12B-IT & LoCoMo & 5.00 & 5.00 & 10.00 & 0.00 & 1,168 & 1,348 \\
\bottomrule
\end{tabular}
\end{table*}

The native recipe therefore usually serializes approximately 20 heterogeneous
objects, not ten flat facts. Small cross-backbone context-length differences
arise because Graphiti extraction builds a different graph for each
generative backbone.

\subsection{LoCoMo Adversarial Category}

The main LoCoMo scores cover categories 1--4, the scope of the released public
runners. At commit \texttt{edcd6f1} the Mem0 harness sets
\texttt{CATEGORIES\_TO\_EVALUATE = [1,2,3,4]}, and the pinned EverMemOS runner
applies \texttt{filter\_category:[5]}~\cite{evermemos}.\footnote{\url{https://github.com/mem0ai/memory-benchmarks/blob/edcd6f1d42400837b1fcb6997716f1769dc51a37/benchmarks/locomo/prompts.py};
\url{https://github.com/EverMind-AI/EverMemOS/blob/539db77d5cc804c875246e34611fd266bf8c1e5d/evaluation/config/datasets/locomo.yaml}}
Category 5 asks whether a question is answerable, and the public prompt assumes
a non-empty reference answer. We therefore report the 446 category-5 questions
separately under the same strict answerability prompt for every method
(Table~\ref{tab:locomo-cat5}). MemoryOS is strongest across all three
backbones; MemForest is competitive with the other retrieval systems but does
not provide a uniform advantage on this category.

\begin{table}[t]
\centering
\small
\caption{LoCoMo category-5 strict answerability pass@1.}
\label{tab:locomo-cat5}
\setlength{\tabcolsep}{3.8pt}
\begin{tabular}{lrrr}
\toprule
Method & Qwen3-4B & Qwen3-30B & Gemma-4-12B-IT \\
\midrule
MemForest & 29.60 & 35.87 & 10.09 \\
EverMemOS & 23.77 & 19.73 & 8.97 \\
LightMem & 11.66 & 14.13 & 14.13 \\
MemoryOS & \textbf{42.38} & \textbf{44.84} & \textbf{28.92} \\
MemPalace & 14.57 & 15.92 & 16.59 \\
Mem0 & 21.30 & 22.42 & 13.90 \\
Zep Local & 10.09 & 11.43 & 6.73 \\
\bottomrule
\end{tabular}
\end{table}
\endgroup

\section{MemForest-Log Scoring Weights}
\label{app:revision-retrieval}
\begingroup

MemForest-Log (Section~\ref{sec:eval-same-pipeline}) scores each fact by
semantic similarity plus weighted lexical and temporal features. Its weights
are selected once on a frozen 82-question development split of mapped LoCoMo
temporal questions, disjoint by question ID from 237 held-out questions, and
then frozen for every evidence budget. The predefined grid is lexical weight
\(\{0,0.04,0.08,0.12\}\) and temporal weight \(\{0,0.5,1.0,1.5\}\).
Table~\ref{tab:log-time-grid} reports development top-10 macro coverage. The
selected pair \((0.12,1.0)\) is the best point in this grid; because 0.12 lies
on the grid boundary, we do not call it a global optimum. Across the 16
settings, development coverage ranges from 76.67\% to 82.60\%.

\begin{table}[H]
\centering
\caption{MemForest-Log development macro coverage (\%).}
\label{tab:log-time-grid}
\setlength{\tabcolsep}{7pt}
\begin{tabular}{r|rrrr}
\toprule
Temporal \textbackslash Lexical & 0 & 0.04 & 0.08 & 0.12 \\
\midrule
0.0 & 77.89 & 77.48 & 79.92 & 81.79 \\
0.5 & 77.89 & 79.11 & 80.33 & 82.20 \\
1.0 & 76.67 & 77.89 & 79.11 & \textbf{82.60} \\
1.5 & 76.67 & 77.48 & 78.70 & 80.57 \\
\bottomrule
\end{tabular}
\end{table}
\endgroup

\section{Fan-Out Sensitivity}
\label{app:design-sensitivity}

\begin{figure}[t]
    \centering
    \includegraphics[width=\columnwidth]{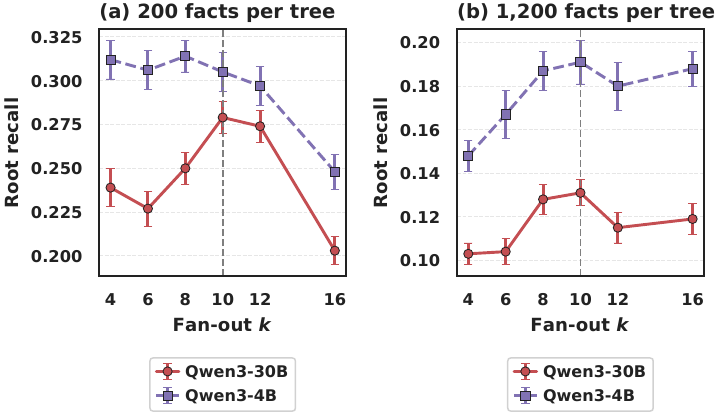}
    \caption{Root recall versus fan-out \(k\) for trees of
(a) 200 and (b) 1,200 facts, with Qwen3-30B and Qwen3-4B summarizers
(\(\pm\)1 SE, 56 histories).}
    \label{fig:memtree-k-sensitivity}
\end{figure}

Figure~\ref{fig:memtree-k-sensitivity} separates the two constraints behind
the default branching factor. Increasing \(k\) makes the tree shallower, but
eventually asks one autoregressive summary call to preserve too many child
facts. The selected moderate operating region balances dependency depth
against summary recall; it is not tuned on the benchmark answer judge.

\section{System Cost, Scaling, Freshness, and Migration}
\label{app:revision-systems}
\label{app:write_path_analysis}
\label{app:migration-scale}
\begingroup
\subsection{Cold- and Warm-Cache Billing}

\begin{table}[t]
\centering
\caption{Cold-start versus verified warm-cache DeepSeek billing. Hit share is
provider-reported hit input divided by total input; cost is US cents per
20-message native probe. Warmups are excluded from both columns.}
\label{tab:deepseek-cold-warm}
\setlength{\tabcolsep}{3.2pt}
\begin{tabular}{lrrrr}
\toprule
Method & Cold hit & Warm hit & Cold cost & Warm cost \\
\midrule
MemForest & 0.00\% & 67.28\% & 0.413 & 0.212 \\
EverMemOS & 1.83\% & 12.06\% & 0.479 & 0.442 \\
Mem0 & 50.60\% & 75.47\% & 0.166 & 0.102 \\
MemoryOS & 7.27\% & 17.52\% & 0.225 & 0.193 \\
Zep Local & 38.60\% & 41.65\% & 1.138 & 1.118 \\
\bottomrule
\end{tabular}
\end{table}

Table~\ref{tab:cache-token-probe-main} reports the verified warm-cache probe
(Appendix~\ref{app:detailed-results}); Table~\ref{tab:deepseek-cold-warm} adds
the cold-start run. The change is largest for MemForest because its long
extraction instruction is shared by one concurrent first wave, and DeepSeek
bills a cache hit only after the prefix has persisted~\cite{deepseek-api-cache}.
We therefore use the warm-cache result for steady-state cost and report the
cold result as a diagnostic.

\subsection{Frozen-Index Query Scaling}

\begin{table}[t]
\centering
\caption{Embedding retrieval as frozen state grows.}
\label{tab:query-scaling}
\begin{tabular}{rrrrrr}
\toprule
Forests & Facts & Trees & Root rows & p50 & p95 \\
\midrule
1 & 1,428 & 250 & 210 & 37.63 ms & 41.79 ms \\
2 & 2,694 & 463 & 369 & 39.14 ms & 43.58 ms \\
4 & 5,553 & 950 & 754 & 39.10 ms & 44.13 ms \\
8 & 10,781 & 1,868 & 1,455 & 39.24 ms & 42.23 ms \\
\bottomrule
\end{tabular}
\end{table}

The sweep uses fixed top-10 embedding browse and excludes answer generation,
index loading, and memory construction. Exact RootIndex recall p95 rises from
0.11 to 0.41 ms as the composed forests grow from 250 to 1,868 trees. A
separate synthetic structural sweep varies both facts per tree and number of
trees. Across the two complete 500-question Qwen builds, the audit covers
250,118 trees; observed maximum height is four and no tree violates the
conservative bounded-fan-out height check.

\subsection{Freshness Contract and Limitations}

One process-local reentrant lock serializes same-user ingest, query, save, and
lifecycle operations. A same-user query blocks during ingest and observes the
rebuilt in-process state after successful return. A source-level probe confirms
that newly committed cells and the refreshed node index become visible
together under this contract and that save and reload preserves the committed
version.

Canonicalization and deduplication do not verify that every extracted fact is
true. An extraction error may propagate into multiple scopes and dirty-node
summaries. Provenance links make the affected leaves and derived artifacts
traceable for deletion and refresh, but automatic fact verification remains
future work.

The implementation does not offer MVCC, concurrent snapshot reads during
refresh, atomic multi-file publication, rollback after partial failure, crash
recovery, multi-process writer safety, or distributed linearizability. These
are limitations rather than inferred guarantees.

\subsection{Migration-State Scaling}

\begin{figure}[t]
    \centering
    \includegraphics[width=\columnwidth]{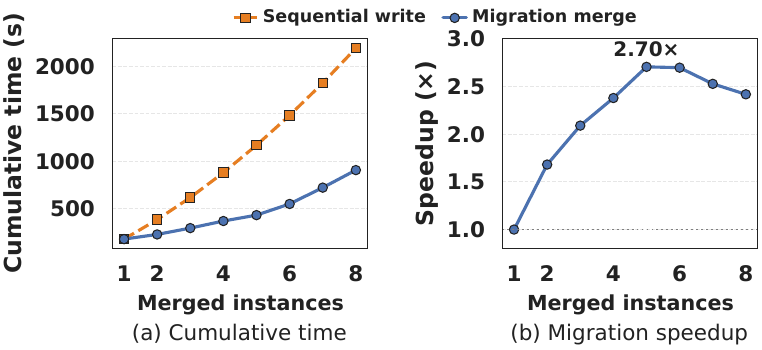}
    \caption{Migration merge versus sequential writes.}
    \label{fig:migration-speedup}
\end{figure}

Figure~\ref{fig:migration-speedup} compares migration merge with sequential
replay of one to eight LongMemEval-S memories. The speedup reaches
\(2.70\times\) at five and six instances and \(2.42\times\) at eight. Because
the instances come from different users, this is a multi-source stress
workload rather than one user's growing history.

Sequential replay and migration merge produce comparable memory states:
facts differ by less than 1\% and trees by at most about 8\%. The procedures
need not be bit-identical because extraction, routing, and summary refresh
contain model or heuristic decisions, but the measured migration speedup does
not come from collapsing or discarding memory state.

\begin{table}[t]
\centering
\caption{Memory scale under sequential write and migration merge.}
\label{tab:app-migration-state}
\setlength{\tabcolsep}{4pt}
\begin{tabular}{rcccc}
\toprule
\(N\) & Seq Facts & Mig Facts & Seq Trees & Mig Trees \\
\midrule
1 & 1,435  & 1,435  & 249  & 249  \\
2 & 2,692  & 2,716  & 375  & 405  \\
3 & 4,061  & 4,098  & 562  & 607  \\
4 & 5,540  & 5,590  & 733  & 792  \\
5 & 6,860  & 6,922  & 895  & 967  \\
6 & 8,101  & 8,174  & 1,044 & 1,128 \\
7 & 9,426  & 9,511  & 1,200 & 1,296 \\
8 & 10,705 & 10,801 & 1,362 & 1,472 \\
\bottomrule
\end{tabular}
\end{table}
\endgroup

\section{Detailed Write-Path and Parallelism Analysis}
\label{app:baseline-write-path-analysis}
\begingroup
This appendix expands Table~\ref{tab:write_path_critical_path}. One complete
memory instance contains \(N\) method-native input units and has a per-instance
LLM worker budget \(P\). LightMem consolidation uses \(P=5\), while MemForest
uses \(P=128\). A star denotes dependency-feasible parallel calls whose
released per-instance ingestion path is serial. Because APIs batch dialogue
differently, \(N\) denotes native units rather than equal-size text chunks;
wall-clock latency is measured on complete question or conversation stores.

The comparison reports autoregressive-LLM dependency depth, not constant-time
token work. Prompt and output length, CPU and database work, serving saturation,
and persistence remain measurable costs. Extraction creates the method's first
stored memory representation; maintenance reconciles, summarizes, or
indexes it. A stage is LLM-based when an autoregressive result lies on its
commit path.

\begin{table*}[t]
\centering
\scriptsize
\caption{Detailed write-path decomposition underlying
Table~\ref{tab:write_path_critical_path}. Complexity denotes logical
dependency depth. A star marks dependency-feasible parallel calls whose
released per-instance ingestion path is serial.}
\label{tab:detailed-write-path-critical-path}
\setlength{\tabcolsep}{3pt}
\renewcommand{\arraystretch}{1.10}
\begin{tabularx}{\textwidth}{@{}
>{\raggedright\arraybackslash}p{0.10\textwidth}
>{\raggedright\arraybackslash}X
>{\centering\arraybackslash}p{0.045\textwidth}
>{\raggedright\arraybackslash}X
>{\centering\arraybackslash}p{0.045\textwidth}@{}}
\toprule
System & Extraction path & LLM & Maintenance path & LLM \\
\midrule
Mem0 & History-independent extraction:
\(O(\lceil N/P\rceil)^*\); released add loop \(O(N)\) & Yes & Candidate
search and reconciliation per add; \(O(N)\) ordered commits & Yes \\
MemoryOS & Raw QA append; \(O(N)\) ordered queue operations & No & At most
\(N\) triggered page, meta-summary, and profile updates, each with up to
\(O(N)\) profile input & Yes \\
EverMemOS & Boundary detection and MemCell formation; \(O(N)\) ordered stream
& Yes & Embedding and index persistence for at most \(N\) emitted cells;
\(O(N)\) work & No \\
LightMem & Post-segmentation extraction:
\(O(\lceil N/P\rceil)^*\); released add loop \(O(N)\) & Yes & Frozen-snapshot
consolidation: \(O(\lceil N/P\rceil)\) independent LLM waves & Yes \\
MemPalace & Deterministic chunk and drawer formation; \(O(N)\) non-LLM work & No &
Embedding and append and index insertion; \(O(N)\) work, no semantic rewrite & No \\
Zep Local & Node and edge extraction per episode; \(O(N)\) ordered episode chain &
Yes & \(O(N)\) ordered LLM chain; backend search \(\sum_i g(G_i)\), index- and
graph-dependent & Yes \\
\textbf{MemForest} & Implemented gather plus semaphore;
\(O(\lceil N/P\rceil)\) request waves & Yes &
\(C_{\mathrm{can}}+\sum_\lambda\lceil|U_\lambda|/P\rceil\) waves;
refresh dependency depth \(O(\log N)\) & Yes \\
\bottomrule
\end{tabularx}
\end{table*}

\subsection{Method-Specific Dependency Boundaries}

\paragraph{Mem0}\label{app:mem0_parallelism}
One add extracts facts, retrieves candidates, and uses a second LLM to choose
add, update, delete, or no-op actions. Isolated extraction prompts are independent,
but the evaluated adapter awaits the complete stateful add before issuing the
next. Concurrent adds can reconcile against and mutate the same old record;
decoupling extraction would require ordered commits, per-record locking, or
optimistic validation. We therefore report the implemented \(O(N)\) add chain
while marking extraction as parallel work.

\paragraph{MemoryOS}\label{app:memoryos_parallelism}
Raw QA append is non-LLM, while queue eviction can trigger page continuity,
meta-summary, profile, and knowledge updates. Profile and knowledge analysis
already use two workers within a trigger, but both depend on selected mid-term
pages and same-user queue and profile commits remain ordered. A hot profile can
also make a triggered LLM input grow with history.

\paragraph{EverMemOS}\label{app:evermemos_parallelism}
Streaming boundary detection observes unresolved conversation history, so
later boundaries depend on earlier decisions and one conversation has
\(O(N)\) ordered extraction depth. Once MemCells are frozen, embedding and
indexing distinct cells add no history-dependent LLM call and can be scheduled
concurrently. Optional foresight and profile extractors are disabled in the
evaluated configuration.

\paragraph{LightMem}\label{app:lightmem_parallelism}
Buffer accumulation and segmentation are ordered, but emitted segments could
be extracted in bounded parallel waves; the released benchmark runner instead
awaits each \texttt{add\_memory}. Its maintenance workers summarize independent
entries from one frozen candidate snapshot and serialize only non-LLM
mutations under a write lock. Thus the evaluated extraction path is serial,
whereas its snapshot-based LLM maintenance already uses \(P=5\) workers.

\paragraph{MemPalace}\label{app:mempalace_parallelism}
Chunk and drawer creation, embedding, and index insertion use no autoregressive
LLM and perform \(O(N)\) work. Preparation and embedding can be
parallelized subject to ordered identifier and timestamp commit. Zero LLM depth
does not imply zero CPU or database cost, and the method performs no LLM
temporal reconciliation.

\paragraph{Zep Local / Graphiti}\label{app:zep_parallelism}
Native \texttt{add\_episode} extracts and resolves nodes and edges, invalidates
superseded facts, and commits graph objects. Episodes in one namespace are
awaited sequentially because later resolution observes the updated graph;
within-episode operations and independent namespaces can still run
concurrently. We denote indexed candidate-search and graph-query work before
episode \(i\) by \(g(G_i)\) because it depends on graph size, indexes, and query
plans. The ordered LLM chain buys versioned facts, provenance, and topology and
is not a claim about every Graphiti deployment.

\paragraph{MemForest}\label{app:memforest_parallelism}
Extraction cells do not read accumulated memory and take
\(\lceil N/P\rceil\) request waves in the batch builder. Canonicalization
contributes \(C_{\mathrm{can}}\) state-dependent waves. Each placement changes
one leaf and \(O(h_s)\) ancestors in scope \(s\); locating the insertion
position is accounted for separately. Refresh issues \(|D|\) summary calls
and schedules independent nodes across trees at each level, giving
\[
\left\lceil N/P\right\rceil+C_{\mathrm{can}}+
\sum_\lambda\left\lceil |U_\lambda|/P\right\rceil
\]
execution waves. The refresh dependency depth is the maximum affected tree
height. Routing, structural insertion, index update, and persistence add
non-LLM work. Realized child-text lengths determine summary token work as
specified in Proposition~\ref{prop:work}.

Overall, the table is a schedule and state-dependency analysis rather than an
impossibility result. Several baselines admit redesigned parallel schedulers;
the reported bounds distinguish those possibilities from the released paths
actually evaluated.

\subsection{MemoryOS Profile Output Cap}
\label{app:profile-cap}

We replay the MemoryOS profile update chain of one LoCoMo history with output
caps of 1k, 2k, 4k, and 8k tokens. Every cap is reached
(Figure~\ref{fig:profile-cap}). A larger cap keeps more facts in the final
view, but each rewrite then regenerates more of the previous version.

\begin{figure}[t]
\centering
\includegraphics[width=\columnwidth]{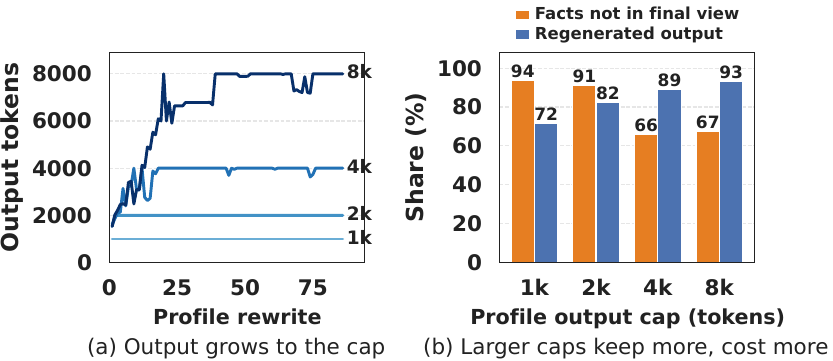}
\caption{MemoryOS profile output cap sweep on one LoCoMo history. (a) Output
tokens per rewrite. (b) Share of facts absent from the final view and share of
each rewrite's output that regenerates the previous version.}
\label{fig:profile-cap}
\end{figure}
\endgroup

\section{Short-Root Evidence-Access Pilot}
\label{app:short-root-pilot}
\begingroup
\begin{figure}[!htbp]
    \centering
    \includegraphics[width=\columnwidth]{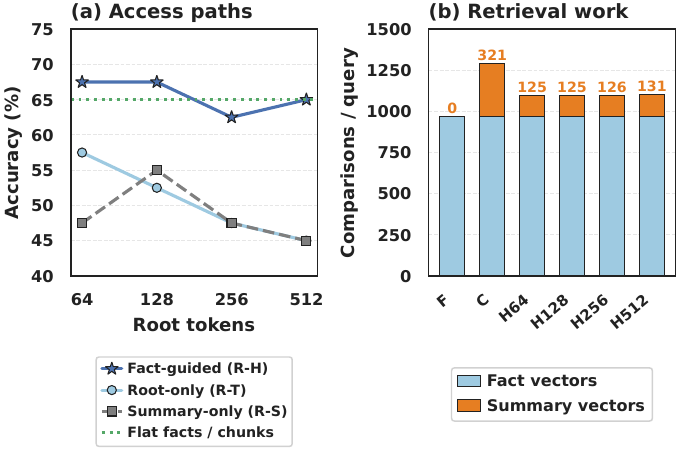}
    \caption{Compact roots with hybrid fact
    access on 40 LoCoMo questions. (a) Accuracy of access paths at four root
    lengths. (b) Vector comparisons.}
    \label{fig:short-root-pilot}
\end{figure}

Figure~\ref{fig:short-root-pilot} gives all access paths at all four root
lengths with fact and summary vector comparisons, on three LoCoMo histories
with a 2,048-token context cap. Flat-fact (F) and indexed-chunk (C) controls
share one line, H is hybrid access, and the realized mean root lengths are 111,
128, 162, and 219 tokens.

\subsection{Frozen Scope}
The pilot selects conv-49, conv-42, and conv-44 by source-turn length
quantiles, excluding development history conv-43. Within the frozen
60-question motivation sample, a seeded hash selects 40 questions:
12 multi-hop, 12 temporal, and 16 single-hop, with 14, 13, and 13 questions in
the selected history order. These histories have been examined previously;
this is not a new unseen test set. No question or root-length setting is
excluded based on its outcome.

The canonical-only states contain 797, 1,008, and 1,117 facts and
95, 114, and 104 roots. They exclude the native session-tree access path.
Of the 313 roots, 210 are L1 nodes and 103 are higher-level nodes.
All four interventions preserve facts, topology, lower summaries, and
their vectors; only root text and its embedding change. The chunk control
keeps the original L1 summaries, including L1 nodes that are also roots.
The root intervention thus leaves the baseline unchanged.

Qwen3-30B-A3B-Instruct-2507-FP8 generates answers and grades
diagnostics, with Qwen3-Embedding-0.6B (1,024 dimensions),
temperature zero, and seed 20261018. Services use vLLM 0.23.0 on H100; the
concurrency ceiling is 128.

\subsection{Access Paths}
R-S ranks all roots by query-vector similarity and packs the top ten
summaries. R-T ranks roots, keeps a width-ten frontier, expands children
until leaves, and scores only facts in the reached leaves. R-H first
scores all facts and roots. Its direct top-30 fact hits nominate containing
roots; the union of top-ten root hits and top-ten nominated roots starts
the width-ten descent. Final candidates include the direct top-30 facts
and facts in the selected leaves. F uses semantic top-500 and lexical
top-50 fact candidates. C selects the ten best original L1 summaries,
unions their facts with the direct top-30 facts, and uses the same final
fact scorer as R-H and F. The scorer combines vector similarity, lexical
overlap with weight 0.12, and fixed query and time features. All arms pack
complete records under the 2,048-token cap, skipping records that do not fit.

This is a common cap, not identical input length: at target 64, R-S uses
937.0 context tokens on average and R-H uses 2,011.5.
Table~\ref{tab:short-root-all} reports every setting, including lower-scoring
hybrid settings and the lower-work R-T control. Vector counts do not
include the whole cost of sorting, lexical matching, data access, or
index loading. They are not GPU or CPU timing measurements. Answers and
scores are cached by exact payload, so the replicated F and C controls are
not additional independent observations.

\begin{table}[t]
\centering
\caption{All short-root controls: correct answers / 40, mean fact and
summary vector comparisons, and mean final context tokens. F and C are invariant
to root length and appear once.}
\label{tab:short-root-all}
\scriptsize
\setlength{\tabcolsep}{3.1pt}
\begin{tabular}{@{}lrrrrr@{}}
\toprule
Arm & Target & Correct & Fact cmp. & View cmp. & Context \\
\midrule
R-S & 64 & 19/40 & 0.0 & 104.1 & 937.0 \\
R-S & 128 & 22/40 & 0.0 & 104.1 & 1020.0 \\
R-S & 256 & 19/40 & 0.0 & 104.1 & 1215.8 \\
R-S & 512 & 18/40 & 0.0 & 104.1 & 1587.5 \\
R-T & 64 & 23/40 & 45.8 & 111.2 & 2012.1 \\
R-T & 128 & 21/40 & 45.3 & 110.2 & 2008.9 \\
R-T & 256 & 19/40 & 45.2 & 110.3 & 2010.8 \\
R-T & 512 & 18/40 & 48.2 & 115.8 & 2014.8 \\
\textbf{R-H} & 64 & 27/40 & 969.6 & 125.2 & 2011.5 \\
\textbf{R-H} & 128 & 27/40 & 969.6 & 125.1 & 2011.5 \\
\textbf{R-H} & 256 & 25/40 & 969.6 & 125.7 & 2011.8 \\
\textbf{R-H} & 512 & 26/40 & 969.6 & 130.7 & 2010.2 \\
F & -- & 26/40 & 969.6 & 0.0 & 2021.0 \\
C & -- & 26/40 & 969.6 & 321.1 & 2011.8 \\
\bottomrule
\end{tabular}

\end{table}

\subsection{Realized Root Lengths}
The generator requests at most 64, 128, 256, or 512 tokens but does not enforce
a hard cap. Table~\ref{tab:short-root-lengths} includes all roots, including
271 failures of the prespecified tolerance $1.25L+16$. Maximum lengths
are 489, 608, 1,183, and 1,289 tokens. Thus the endpoint intervention reduces
actual root text by 49.3\%, rather than the eightfold ratio of requested
targets; no output was truncated to manufacture compliance.

\begin{table}[t]
\centering
\caption{Realized root lengths (tokens), 313 roots per target $L$.
Last columns count roots beyond the target and its tolerance.}
\label{tab:short-root-lengths}
\scriptsize
\setlength{\tabcolsep}{3.2pt}
\begin{tabular}{@{}rrrrrr@{}}
\toprule
$L$ & Mean & Median & P90 & $>L$ & $>1.25L+16$ \\
\midrule
64 & 111.2 & 83 & 217.8 & 196 & 132 \\
128 & 127.5 & 92 & 274.6 & 116 & 83 \\
256 & 161.9 & 97 & 360.4 & 63 & 38 \\
512 & 219.5 & 112 & 516.0 & 32 & 18 \\
\bottomrule
\end{tabular}

\end{table}

\subsection{Interpretation Limits}
The overall advantage does not imply dominance on every history or question
type: at target 64, R-H answers 5/12 temporal questions against 7/12 for R-T,
but 14/16 single-hop questions against 9/16. R-H and C have identical final
contexts for 32/40 questions, and R-H and F for 20/40; 98.75\% of R-H's
selected fact occurrences belong to the common direct top-30 candidates. The
intervention therefore supports combined fact access, not a claim that
fact-to-tree nomination alone explains the gain.

All accuracy values use the frozen same-model diagnostic judge. A spot check
finds one suspected false positive (R-S at target 64 for
\texttt{conv-49\_qa062}), kept without selective rescoring. The pilot
establishes neither a population-level quality advantage over F and C nor
reductions in write tokens or physical query latency.
\endgroup

\clearpage

\end{document}